\documentclass[aps,preprintnumbers,amsmath,amssymb,prd,superscriptaddress,longbibliography,twocolumn]{revtex4-2}
\usepackage{dcolumn}
\usepackage{color}
\usepackage[caption=false]{subfig}
\usepackage{feynmp}
\usepackage{feynmp-auto}
\usepackage{float}
\usepackage{bbm}
\usepackage{empheq}

\def\comment#1{}

\newcommand{\nc}{\newcommand}
\nc{\beq}{\begin{eqnarray}}
	\nc{\eeq}{\end{eqnarray}}
\nc{\scs}{\scriptstyle}
\nc{\setval}{\fmfset{wiggly_len}{3mm} \fmfset{arrow_len}{1.5mm}
	\fmfset{arrow_ang}{13} \fmfset{dash_len}{1.5mm}\fmfpen{0.125mm}
	\fmfset{dot_size}{2thick}}

\usepackage{bm,latexsym,mathrsfs,enumerate,color}
\usepackage[mathcal]{euscript}
\usepackage[breaklinks=true,unicode=true,urlcolor = blue,colorlinks = true,citecolor = blue,linkcolor = blue]{hyperref}
\usepackage{graphicx}
\usepackage{todonotes}
\usepackage{wrapfig}

\def\slashchar#1{\setbox0=\hbox{$#1$}           
	\dimen0=\wd0                                 
	\setbox1=\hbox{/} \dimen1=\wd1               
	\ifdim\dimen0>\dimen1                        
	\rlap{\hbox to \dimen0{\hfil/\hfil}}      
	#1                                        
	\else                                        
	\rlap{\hbox to \dimen1{\hfil$#1$\hfil}}   
	/                                         
	\fi}                                         %

\DeclareMathAlphabet\mathbfcal{OMS}{cmsy}{b}{n}

\usepackage{physics}
\usepackage{xcolor}

\begin{document}



\title{Walking behavior induced by $\mathcal{PT}$ symmetry breaking in a non-Hermitian $\rm XY$ model with clock anisotropy}

\author{Eduard Naichuk}
\affiliation{Institute for Theoretical Solid State Physics, IFW Dresden, Helmholtzstr. 20, 01069 Dresden, Germany}
\affiliation{Bogolyubov Institute for Theoretical Physics, 03143 Kyiv, Ukraine}

\author{Jeroen van den Brink}
\affiliation{Institute for Theoretical Solid State Physics, IFW Dresden, Helmholtzstr. 20, 01069 Dresden, Germany}
\affiliation{Institute for Theoretical Physics and W\"urzburg-Dresden Cluster of Excellence ct.qmat, TU Dresden, 01069 Dresden, Germany}

\author{Flavio S. Nogueira}
\affiliation{Institute for Theoretical Solid State Physics, IFW Dresden, Helmholtzstr. 20, 01069 Dresden, Germany}

\date{\today}

\begin{abstract}
A quantum system governed by a non-Hermitian Hamiltonian may exhibit zero temperature phase transitions that are driven by interactions, just as its Hermitian counterpart, raising the fundamental question how non-Hermiticity affects quantum criticality. In this context we consider a non-Hermitian system consisting of an $\rm XY$ model with a complex-valued four-state clock interaction that may or may not have parity-time-reversal ($\mathcal{PT}$) symmetry. When the $\mathcal{PT}$ symmetry  is broken, and time-evolution becomes non-unitary, a scaling behavior similar to the Berezinskii-Kosterlitz-Thouless 
phase transition ensues, but in a highly unconventional way, as the line of fixed points is absent. From the analysis of the $d$-dimensional RG equations, we obtain that the unconventional behavior in the $\mathcal{PT}$ broken regime follows from the collision of two fixed points in the $d\to 2$ limit, leading to walking behavior or pseudocriticality. For $d=2+1$ the near critical behavior is characterized by a correlation length exponent $\nu=3/8$, a value smaller than the mean-field one. 
These results are in sharp contrast with the $\mathcal{PT}$-symmetric case where only one fixed point arises for $2<d<4$ and in $d=1+1$ three lines of fixed points occur with a continuously varying critical exponent $\nu$.  
\end{abstract}

\maketitle


\section{Introduction}
The possibilities to access non-Hermitian systems \cite{Bender_PhysRevLett.80.5243,Bender-Introduction,Ashida-Review} experimentally significantly expanded in the past years, with possible incarnations ranging from metamaterials to ultracold atoms realizations~\cite{Thomale_PhysRevLett.126.215302,Metamaterial,Ultracold_PhysRevLett.129.070401}, and more recently including electronic systems having applications as sensors~\cite{Kiril,koenye2023nonhermitian,Libo_Ma}. 
While in any realistic physical system interactions play an important role, from a theoretical perspective {\it interacting} non-Hermitian systems have been little explored. Particularly pertinent is the question how non-Hermiticity affects interaction-driven quantum phase transitions and their universal physical properties also in absence of $\mathcal{PT}$  symmetry, where time-evolution becomes non-unitary \cite{Hatano-Nelson_PhysRevLett.77.570,BENDER2005333,article,Ashida-Review,Neupert_PhysRevB.106.L121102}. This motivates considering quantum field theoretic models, where the renormalization group (RG) provides the defining paradigm for exploration of the phase structure. 
In this context it is very important to establish how the critical behavior depends on the spacetime dimensionality $d=D+1$.

In order to address the questions above and understand how the breaking of $\mathcal{PT}$ symmetry affects the phase structure, we investigate an effective quantum field theory Lagrangian consisting of an $\rm XY$ model with a non-Hermitian, $\mathcal{PT}$-symmetric clock interaction in $d=D+1$ spacetime dimensions,
{\color{black}
\begin{equation}
	\label{Eq:Leff-0}
	\mathcal{L}=\frac{K}{2}(\partial_\mu\theta)^2-z_r\cos N\theta-iz_i\sin N\theta.
\end{equation}
The $\mathcal{PT}$ symmetry is realized in the form, $\mathcal{P}\theta\mathcal{P}^{-1}=-\theta$ and $\mathcal{T}i\mathcal{T}^{-1}=-i$.  	
}
Its Hermitian counterpart ($z_i=0$) arises in several contexts both in high-energy physics \cite{THOOFT19781,Elitzur_PhysRevD.19.3698} and condensed matter physics \cite{Jose_PhysRevB.16.1217,KADANOFF197939,Hove-Sudbo_PhysRevE.68.046107,Oshikawa_PhysRevB.61.3430,Damle_PhysRevB.91.104411,Sandvik_PhysRevLett.124.080602,Sandvik_PhysRevB.103.054418}.  
{\color{black} If the periodicity of the scalar field is ignored, the clock model of Eq. (\ref{Eq:Leff-0}) just becomes the non-Hermitian sine-Gordon theory considered earlier within a bosonization duality context in Ref. \cite{BENDER2005333} and whose quantum critical behavior has been analyzed in Ref. \cite{article}. In this case the value of $N$ is immaterial and can easily be scaled out from the problem. However, some features of the sine-Gordon system still matter when $\theta$ is dealt as periodic scalar field \cite{Jose_PhysRevB.16.1217}. Importantly, a similarity transformation discussed in Ref. \cite{BENDER2005333} allows to map the Lagrangian above into a Hermitian one,
	\begin{equation}
		\label{Eq:Leff-0-Hermitian}
		\mathcal{L}'=\frac{K}{2}(\partial_\mu\theta)^2-\sqrt{z_r^2-z_i^2}\cos N\theta.
	\end{equation} 
	For a non-periodic $\theta$ and $z_r<z_i$, the Lagrangian (\ref{Eq:Leff-0-Hermitian}) becomes a so called ''imaginary sine-Gordon" (ISG) theory, whose mathematical-physical properties were extensively studied in the past \cite{ZAMOLODCHIKOV1994436,Fendley,Castro-Alvaredo_2017}. As such it is an example of non-unitary quantum field theory which has applications to polymer physics \cite{Fendley}. Viewed as a transformation from the $\mathcal{PT}$-symmetric Lagrangian (\ref{Eq:Leff-0}), the ISG corresponds to a regime where the $\mathcal{PT}$ symmetry is broken \footnote{Note that we could equally consider the ISG as being $\mathcal{PT}$-symmetric with $\mathcal{P}\theta\mathcal{P}^{-1}=\theta+\pi/N$, but then the $\mathcal{PT}$ symmetry would be broken in the Lagrangian (\ref{Eq:Leff-0}).}. 

As we will see in detail in this paper, the periodicity of $\theta$ underlying the non-Hermitian clock model significantly affects the phase structure relative to the non-Hermitian sine-Gordon behavior.} 
From a renormalization group (RG) perspective the $N$-state clock interaction is an irrelevant operator for $N> 4$ in spacetime dimensions $2<d<4$, so one expects a crossover behavior from the $\rm XY$ universality class to $Z_N$ symmetric models. In particular, the case $N=4$ corresponds in the strong coupling clock interaction limit to two decoupled Ising regimes \cite{Damle_PhysRevB.91.104411,Sandvik_PhysRevLett.124.080602}.   

Since the $\mathcal{PT}$-symmetric clock model can be mapped into its Hermitian counterpart, the fundamental question to answer is how $\mathcal{PT}$ {\it symmetry breaking} changes the phase structure of the system in $d=D+1$ dimensions. In particular, we would like to establish whether a well defined phase transition occurs.

Here we show that when the $\mathcal{PT}$ symmetry is broken for $2<d<4$ the RG flow is characterized by two fixed points rather than only one in the $\mathcal{PT}$-symmetric case, and that near one of them complex conjugated scaling dimensions emerge. This leads to a phase transition in $d=2+1$ dimensions, with a near critical behavior characterized by a correlation length exponent $\nu=3/8$. As $d\to 2$ the two fixed points collide, leading to a so called 'walking' behavior \cite{Rychkov-Walking}, sometimes also referred to as pseudocriticality, a concept that has been in focus in recent years both in high-energy and condensed matter physics \cite{Rychkov-Walking,Son_PhysRevD.80.125005,Nogueira_2013,Nahum_PhysRevX.5.041048,Nogueira_PhysRevD.100.085005,Ma-Wang_PhysRevB.102.020407,Scherer_PhysRevB.100.134507,Ma-He_PhysRevB.99.195130,Burgelman_PhysRevE.107.014117,Weber-Vojta_PhysRevLett.130.186701,hawashin2023nordicwalking,weber2024tunable,zhou2024mathrmso5,jacobsen2024lattice,tang2024reclaiming}. The resulting  scaling behavior for $d\to 2$ is similar to a   Berenzinskii-Kosterlitz-Thouless (BKT) one. However, the actual phase transition is not a BKT one, since the RG flow features a fixed point rather than a line of fixed points. 
We obtain that the flow lines walk around the fixed point on RG-invariant surfaces. 

{\color{black} Simply stated \cite{Son_PhysRevD.80.125005,Rychkov-Walking}, walking can be exemplified by the following toy RG $\beta$ function,
\begin{equation}
	\beta(g)=\mu\frac{dg}{d\mu}=\alpha-\alpha_*-(g-g_*)^2,
\end{equation}
where $\mu$ is a scale variable and $\alpha$ is a parameter that is protected and not being affected by scale transformations. The RG flow is then characterized by two fixed points,
\begin{equation}
	g_\pm=g_*\pm\sqrt{\alpha-\alpha_*}, 
\end{equation} 
with $g_-=\lim_{\mu\to 0}g(\mu)$ and $g_+=\lim_{\mu\to \infty}g(\mu)$ corresponding to infrared (IR) and (UV) fixed points. Trivially, these fixed points are real for $\alpha>\alpha_*$. Things become interesting when these fixed points collide as $\alpha\to\alpha_*$. After this fixed point collision, decreasing $\alpha$ further turn $g_\pm$ complex. In this case solving the RG equation leads to the the following ratio between IR and UV scales,
\begin{equation}
	\frac{\mu_{\rm IR}}{\mu_{\rm UV}}\sim \exp\left(-\frac{\rm const}{\sqrt{\alpha_*-\alpha}}\right).
\end{equation}
This type of behavior is representative of a walking scaling \cite{Rychkov-Walking}. Although the above equation has a form reminiscent of a BKT scaling, it does not actually completely abide to a BKT type of behavior, as discussed in Ref. \cite{Rychkov-Walking}. This will also be the point in our analysis of the $\mathcal{PT}$ symmetry broken state of the four-state clock model. 

The plan of this paper is as follows. In order to introduce the non-Hermitian model considered here we first consider a toy classical model in Section II that corresponds to a zero-dimensional version (in the sense that no derivatives are involved) of the non-Hermitian clock model. This case is easily solvable exactly and serves as a starting point to the basic ideas put forward in this Introduction. Despite having a complex Hamiltonian of the form exhibited by the potential of Eq. (\ref{Eq:Leff-0}), the partition function obtained in Section II is always real. Section III considers a quantum mechanical non-Hermitian model with $Z_N$ anisotropy and we explicitly construct the similarity transformation to a Hermitian model in the canonical quantzation formalism. This example also helps to illustrate how one can view gain and loss in terms of particle creation and annihilation. Section IV discusses the quantum field theory non-Hermitian model already introduced above and in Section V the RG analysis is performed and the walking behavior is derived. Section VI concludes the paper and Appendices A and B give further details of the RG analysis.     
}

{\color{black}
\section{A toy classical model}

In order to illustrate the main ideas contained in this paper at a very basic level, let us first consider the classical partition function of a toy model with a complex Hamiltonian, 
\begin{equation}
	\label{Eq:Toy-H}
H=-J\cos\theta-iK\sin\theta,
\end{equation} 
where $\theta\in[0,2\pi]$ and $J,K>0$. Since a time variable does not play a role in this system, the role of time-reversal is simply played by complex conjugation and we define the parity transformation as $\theta\to -\theta$, as it corresponds to a reflection from  a point on the unit circle $x^2+y^2=1$ on an antipodal point. The Hamiltonian of Eq.~(\ref{Eq:Toy-H}) is clearly invariant under the operation of complex conjugation followed by a parity transformation, which corresponds in this case to a toy version of the quantum $\mathcal{PT}$-symmetry. Even more specifically, it corresponds to zero-dimensional version of the non-Hermitian, $\mathcal{PT}$-symmetric sine-Gordon theory considered in Ref.~\cite{BENDER2005333}. 

The partition function, 
\begin{equation}
	\label{Eq:Z-toy}
	Z=\int_{0}^{2\pi}\frac{d\theta}{2\pi}e^{\beta(J\cos\theta+iK\sin\theta)}, 
\end{equation} 
is readily evaluated exactly,
\begin{equation}
	\label{Eq:Z-toy-1}
	Z= I_0(\beta\sqrt{J^2-K^2}),
\end{equation}
where $I_0(x)$ is a modified Bessel function of the first kind. We see that despite having a complex Hamiltonian, the partition function is well defined for all values of $J$ and $K$. The result is a consequence of the fact that the toy Hamiltonian of Eq.~(\ref{Eq:Toy-H}) can be mapped into an equivalent, real Hamiltonian, 
\begin{equation}
	\label{Eq:Toy-H-1}
	H'=-\sqrt{J^2-K^2}\cos\theta.
\end{equation} 

What we have illustrated in this toy example is precisely what happens in non-Hermitian systems with $\mathcal{PT}$ symmetry: they can be mapped into an Hermitian system by a similarity transformation \cite{Bender-Introduction,Ashida-Review}. Performing such a transformation is not always easy, but an example as straightforward as in our toy model has been considered earlier, namely, the non-Hermitian $\mathcal{PT}$-symmetric sine-Gordon model in 1+1 dimensions \cite{BENDER2005333}, featuring a complex sine-Gordon potential having the same form as the one in Eq.~(\ref{Eq:Toy-H}). 

Note that the free energy vanishes for $J=K$, something that is immediately obvious by looking at the Hamiltonian $H'$ above, but this is not apparent from the original Hamiltonian of Eq.~(\ref{Eq:Toy-H}), which becomes $H=-Je^{i\theta}$. In an actual quantum field theory realm, this would correspond to the form of a so called time-like Liouville theory (recall that in a Liouville theory the potential has the form, $V(\phi)=-\mu e^{-b\phi}$) \cite{Saleur_PhysRevLett.116.130601}. However, both in the sine-Gordon and Liouville theories the scalar field is not a periodic field, unlike the case of the toy example considered here. A sine-Gordon model featuring a periodic scalar field is actually a clock model or, more precisely, an $\rm XY$ model with clock anisotropy. Importantly, in the actual field theory replacing $\cos\theta\to\cos N\theta$, $N\in\mathbb{N}$ makes a significant difference depending on the value of $N$ \cite{Jose_PhysRevB.16.1217}, while in our toy example the value of $N$ is immaterial. 

\section{Quantum mechanical model}

We have mentioned in the previous section that our classical toy model features a periodic scalar field and that we are going to consider a quantum field theory with the same potential later on, i.e., it is not just a non-Hermitian sine-Gordon theory, but rather an $\rm XY$ model with clock anisotropy. Before discussing the actual quantum field theory, let us first discuss the quantum mechanical problem. At this stage it is interesting to recall the well known fact that it is technically challenging to consider the phase variable $\theta$ as an Hermitian operator, a difficulty that has been a subject of controversy spanning several decades \cite{Carruthers-Nieto_RevModPhys.40.411,DTPegg_1988}. This fact, while important, is not acting significantly in the type of analysis we make in this paper. Here we take mostly a path integral point of view, which features the classical action, where the definition of a phase variable is less problematic, although here too other important technical aspects come into play, in view of the functional integration measure containing a periodic field. Nevertheless, within the path integral formalism at least the semi-classical picture of the problem comes out right. With these preparatory remarks in mind, let us consider a quantum mechanical bosonic problem that can be considered as a version of a non-Hermitian clock model, 
\begin{equation}
	\label{Eq:H-clock}
	H=\epsilon a^\dagger a-\frac{1}{2}\left[(J+K)a^N+(J-K)(a^\dagger)^N\right],
\end{equation}         
where $a$ and $a^\dagger$ are the usual bosonic annihilation and creation operators. 
Since $\mathcal{PT}$ symmetry maps both $a$ and $a^\dagger$ into $-a$ and $-a^\dagger$, the Hamiltonian is $\mathcal{PT}$-symmetric only if $N$ is even. 

The expectation value of the Hamiltonian (\ref{Eq:H-clock}) with respect to the standard $U(1)$ coherent state basis, $|\alpha\rangle$ yields the semi-classical complex Hamiltonian, $\mathcal{H}=\langle\alpha|H|\alpha\rangle$,  
\begin{equation}
	\label{Eq:H-clock-semi-classical}
	\mathcal{H}=\epsilon |\alpha|^2-|\alpha|^N(J\cos N\theta+iK\sin N\theta),
\end{equation}
where $\theta$ is defined via $\alpha=|\alpha|e^{i\theta}$. The semi-classical Hamiltonian has a form similar to the one exhibited by the toy model of the previous section. 

Let us first assume that $J>K$, so that the process of creating or annihilating $N$ particles are both associated to some specified negative energy amplitudes. When the above Hamiltonian acts on a state $|n\rangle$ having $n$ particles, creating additional $N$ particles would cost less energy than annihilating $N$ of them. Thus, the physical process that follows from applying the Hamiltonian (\ref{Eq:H-clock}) to the state $|n\rangle$ does not conserve energy. This is one instance of a typical gain and loss process that is characteristic of many non-Hermitian systems \cite{Ashida-Review}. Nevertheless, the energy spectrum is real, provided we keep $J>K$. To see this, let us consider the similarity transformation, 
\begin{equation}
	H'=e^{\lambda a^\dagger a}He^{-\lambda a^\dagger a},
\end{equation}
where $\lambda=N^{-1}{\rm arctanh}(K/J)$. This yields, 
\begin{equation}
	\label{Eq:H-prime}
	H'=\epsilon a^\dagger a -\frac{\sqrt{J^2-K^2}}{2}\left[a^N+(a^\dagger)^N\right],
\end{equation}
which is indeed Hermitian for $J>K$. In this form the $\mathcal{PT}$-symmetry breaking for $J<K$ is apparent. Not only the Hamiltonian would become non-Hermitian once more, but it would fail to be  $\mathcal{PT}$-symmetric for even $N$. However, now it would be $\mathcal{PT}$-symmetric for $N$ odd, which was not the case for $J>K$. This $\mathcal{PT}$ symmetry for $N$ odd is reminiscent of the $\mathcal{PT}$ symmetry in the Lee model \cite{Bender_PhysRevD.71.025014}, which feature a vertex with three particles and an imaginary coupling constant in its non-Hermitian regime.

Let us give another quantum mechanical example based on a double-well Bose-Einstein condensate system that allows the implementation of a Josephson tunneling via population imbalance across the wells \cite{Oberthaler_2007}. We can model this in the form of a two-site Bose-Hubbard model \cite{Bose-Hubbard_PhysRevB.40.546} with Hamiltonian,
\begin{equation}
	\label{Eq:2-site-BH}
	H=-J(a_1^\dagger a_2+a_2^\dagger a_1)+\sum_{i=1,2}\left[-\mu n_i+\frac{U}{2}n_i(n_i-1)\right],
\end{equation} 
where $n_i=a_i^\dagger a_i$ is the particle number operator at well $i$ and $\mu$ is the chemical potential. Both the tunneling amplitude $J$ and the bosonic Hubbard interaction $U$ are positive. The semi-classical Hamiltonian associated to the Josephson tunneling considered in the review article \cite{Oberthaler_2007} is obtained by taking the expectation value of the Hamiltonian (\ref{Eq:2-site-BH}) in the coherent state, $|\Psi\rangle=|\alpha_1,\alpha_2\rangle$, satisfying, $a_i|\Psi\rangle=\alpha_i|\Psi\rangle$, where $\alpha_i$ are complex numbers such that $|\alpha_i|^2=\bar n_i$ yields the average particle number at the well $i$. In this way we obtain $\mathcal{H}=\langle\Psi|H|\Psi\rangle$ in the form,
\begin{equation}
	\label{Eq:Josephson-BEC}
	\mathcal{H}=E_n-J\sqrt{n^2-(\Delta\bar n)^2}\cos\Delta\theta+\frac{U}{4}(\Delta\bar n)^2,
\end{equation}
where $E_n=-(\mu+U/2)n+Un^2/4$, $n=\bar n_1+\bar n_2$ is the total particle number in the system, $\Delta\bar n=\bar n_1-\bar n_2$, and $\Delta\theta=\theta_1-\theta_2$ is the phase difference across the wells. The phases are defined via $\alpha_i=\sqrt{\bar n_i}e^{i\theta_i}$. As a closed system we always have $n>\Delta\bar n$. However, for an open system we can consider $\bar n_1=\bar n_{10}+\delta\bar n$ and $\bar n_2=\bar n_{20}-\delta\bar n$ corresponding to a situation of gain and loss in the wells 1 and 2, respectively. This leaves $n$ invariant, but $\Delta\bar n=\bar n_{10}-\bar n_{20}+2\delta\bar n$. Hence, if $\delta\bar n>\bar n_{20}$ we obtain that the Josephson tunneling term becomes purely imaginary, in a situation characteristic of a system with broken $\mathcal{PT}$ symmetry. 

In order to change the periodicity in this Josephson tunneling example we can potentially consider a solid state device rather than a Bose-Einstein system. Twisted multi-layer systems may be employed in this case \cite{Jarillo-Herrero-NatMat,Poccia-Kim_Science}. 
}

\section{Quantum field theory}

Let us consider a non-Hermitian theory consisting of a $U(1)$-invariant Lagrangian with a $Z_{N}$ anisotropy perturbation,
\begin{eqnarray}
	\label{NHLagrangian}
    \mathcal{L}_{\text{NH}}&=&|\partial_\mu\psi|^2+m^2|\psi|^2+\frac{u}{2}|\psi|^4
\nonumber\\
&-&\frac{1}{2}\left[(v+w)\psi^N+(v-w){\psi^*}^N\right],
\end{eqnarray}
where $u, v, w>0$. The $Z_N$ anisotropic interaction is irrelevant in the RG sense for $N>4$ when the spacetime dimension lies in the interval $2<d<4$. For $w=0$ this expectation is corroborated by recent Monte Carlo simulations \cite{Sandvik_PhysRevLett.124.080602}. Our main focus here will be on the $N=4$ case. For $d=2$ a special treatment is required, which is one of the main topics in the subsequent discussion. The Hermitian model for $N=4$ has been studied in different contexts in the past decades. For example, it arises as an effective theory emerging from breaking the $SU(N)$ symmetry in a non-Abelian Higgs model \cite{THOOFT19781} and more recently for $N=4$ and $d=3$  as theories describing the transition to valence-bond solid phases \cite{Damle_PhysRevB.91.104411}.    

The connection to the $N$-state clock model is easily established after employing the following parametrization of the scalar field, $\psi=\rho e^{i\theta}/\sqrt{2}$, in which case the Lagrangian becomes,
\begin{eqnarray}
	\label{NHLagrangian-1}
	\mathcal{L}_{\text{NH}}&=&\frac{1}{2}(\partial_\mu\rho)^2+\frac{\rho^2}{2}(\partial_\mu\theta)^2+\frac{m^2}{2}\rho^2+\frac{u}{8}\rho^4
	\nonumber\\
	&-&\frac{\rho^N}{2^{N/2}}\left[v\cos(N\theta)+iw\sin(N\theta)\right]. 
\end{eqnarray} 
In the representation above we see that this non-Hermitian theory is $\mathcal{PT}$-symmetric, where $\mathcal{P}\theta\mathcal{P}^{-1}=-\theta$ and $\mathcal{T}i\mathcal{T}^{-1}=-i$. The amplitude $\rho$ is both $\mathcal{P}$- and $\mathcal{T}$-invariant. 

Near criticality phase fluctuations are dominant, so we can consider the amplitude field $\rho$ as being essentially frozen, and we can approximately write $\rho^2=\rho_0^2=-2m^2/u$ for $m^2<0$. In this case we obtain the effective clock model, 
\begin{equation}
	\label{Eq:Leff-1}
	\mathcal{L}_{\rm eff}=\frac{K}{2}(\partial_\mu\theta)^2-z_r\cos N\theta-iz_i\sin N\theta,
\end{equation} 
where $K=\rho_0^2$ (this parameter $K$ should not be confused with the coupling $K$ used in the previous two sections), $z_r=v\rho_0^N/2^{N/2}$, and  $z_i=w\rho_0^N/2^{N/2}$. For $z_i=0$ the model is well studied in the literature, where in particular Jos\'e {\it et al.} \cite{Jose_PhysRevB.16.1217} considered the two-dimensional $\rm XY$ model with clock-type symmetry breaking fields.

\section{Renormalization group analysis}

A major difference with respect to the theories discussed in Refs.~\cite{BENDER2005333} and \cite{article} lies in the fact that the scalar field $\theta$ is a periodic field. Hence, for $z_i=0$ the Lagrangian above yields the effective theory for an $\rm XY$ model with a $N$-fold clock symmetry breaking field \cite{Jose_PhysRevB.16.1217}. 

Within a path integral formalism we can perform a shift of the field, $\theta\to\theta+(i/N){\rm arctanh}(z_i/z_r)$ \cite{BENDER2005333}. This leads us to consider the following equivalent effective Lagrangian, 
\begin{equation}
	\label{Eq:Leff-2}
	\mathcal{L}_{\rm eff}=\frac{K}{2}(\partial_\mu\theta)^2-\widetilde{z}\cos N\theta,
\end{equation}
where $\widetilde{z}=\sqrt{z_r^2-z_i^2}$. When the $\mathcal{PT}$-symmetry is broken, we simply have $\widetilde{z}\to i\widetilde{z}$. Thus, we see that the couplings $z_r$ and $z_i$ are not actually independent. This leads to the non-renormalization of the ratio $z_r/z_i$, which ultimately is what will lead to the walking behavior in the  $\mathcal{PT}$ broken regime. 

We will follow a procedure similar to the one by Jos\'e {\it et al.}~\cite{Jose_PhysRevB.16.1217} in their analysis of the $\rm XY$ model in the presence of symmetry breaking fields (the simplest case refers to an external in-plane magnetic field). First, we note that when $\widetilde{z}=0$ the effective Lagrangian (\ref{Eq:Leff-2}) for $d=2$ is exactly dual to a sine Gordon theory of the form \cite{Jose_PhysRevB.16.1217,frohlich1981statistical},
\begin{equation}
	\mathcal{L}_{\rm SG}=\frac{1}{8\pi^2K}(\partial_\mu\chi)^2-z\cos\chi,
\end{equation} 
corresponding to the field theory of a Coulomb gas of point vortices \cite{frohlich1981statistical}. Hence, defining the dimensionless renormalized couplings $y$ and $\kappa$ associated respectively to the bare couplings $z$ and $K$, we obtain the usual RG equations underlying the BKT transition \cite{Jose_PhysRevB.16.1217,Nelson-Kosterlitz1977}. 

From now on we focus on the $N=4$ and discuss the clock model symmetry breaking perturbation for this case. Let us define the dimensionless couplings $y_r$ and $y_i$ associated to $z_r$ and $z_i$, respectively. The RG equations in $d$ dimensions are given by,
\begin{equation}
	\label{Eq:dk-final-maintext}
	\frac{d\kappa}{dl}=\frac{4}{\pi^2}(y_r^2-y_i^2)-\kappa^2y^2+(d-2)\kappa,
\end{equation}
\begin{equation}
	\label{Eq:dy-maintext}
	\frac{dy}{dl}=\left[d-f(d)\kappa\right]y. 
\end{equation}
\begin{equation}
	\label{Eq:dyr-maintext}
	\frac{dy_r}{dl}=\left[d-\frac{4f(d)}{\pi^2\kappa}\right]y_r,
\end{equation}  
\begin{equation}
	\label{Eq:dyi-maintext}
	\frac{dy_i}{dl}=\left[d-\frac{4f(d)}{\pi^2\kappa}\right]y_i,
\end{equation}
where $f(d)=(d-2)\Gamma(d/2-1)/(2\pi^{d/2-2})$. These computations are given in the Appendix A.

\begin{figure}
	\subfloat[$d=3$ $\mathcal{PT}$ symmetry can be broken or not, $\widetilde{y}=0$]{\includegraphics[width=0.5\linewidth]{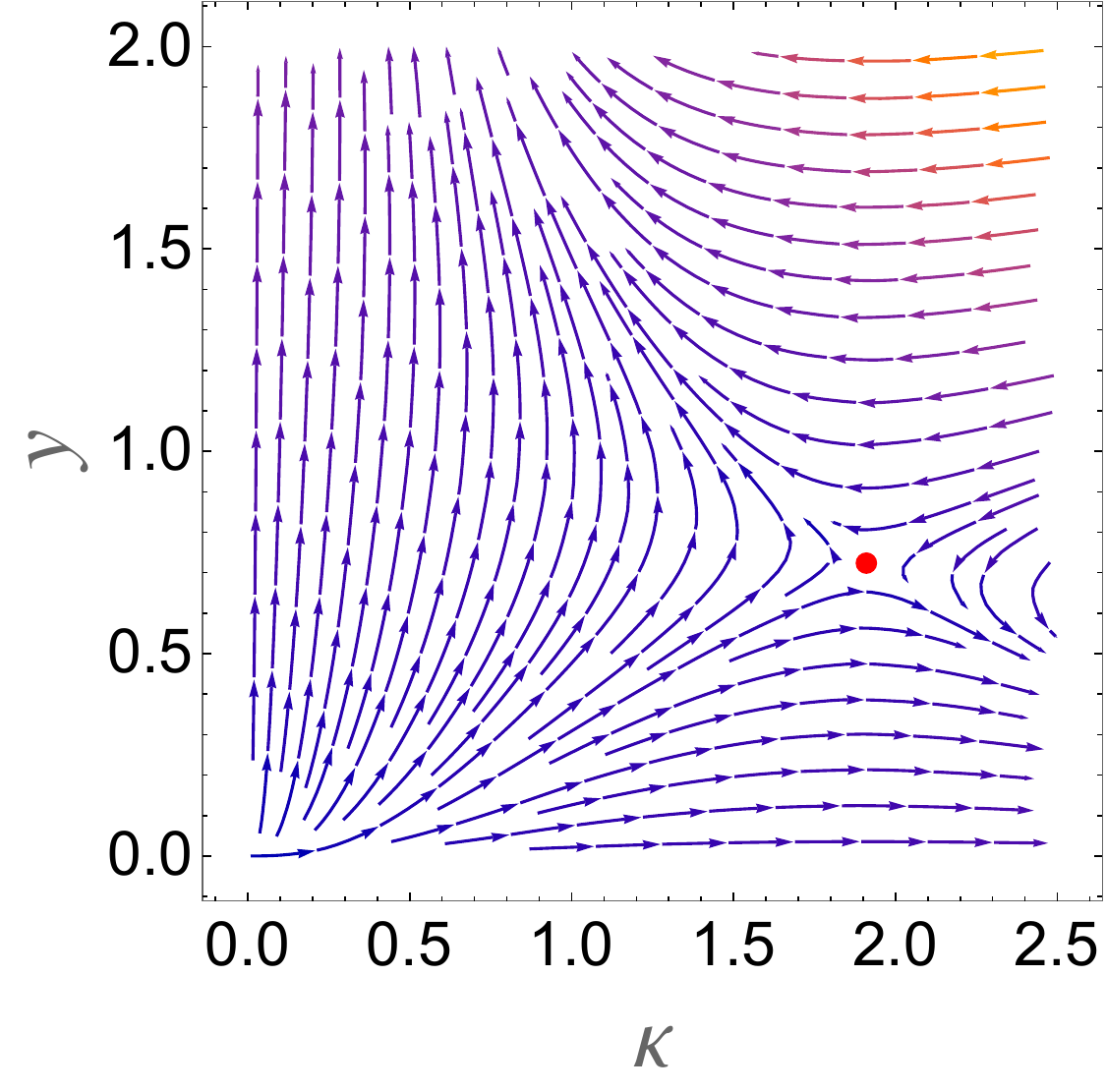}}
	\hfill
	\subfloat[$d=3$ $\mathcal{PT}$ broken, $y=0$]{\includegraphics[width=0.5\linewidth]{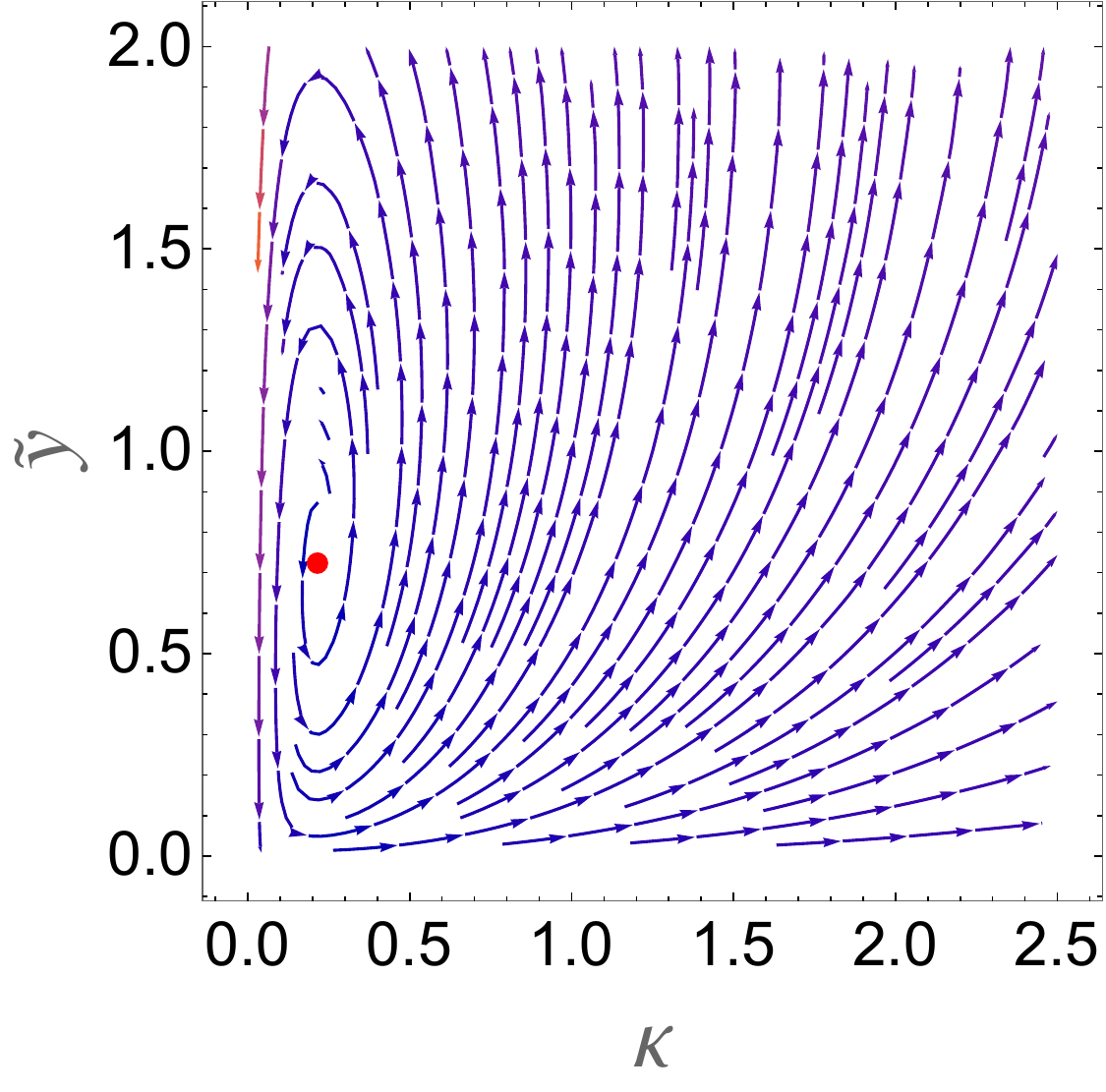}}\\
	\subfloat[$d=3$ $\mathcal{PT}$-symmetric]{\includegraphics[width=0.5\linewidth]{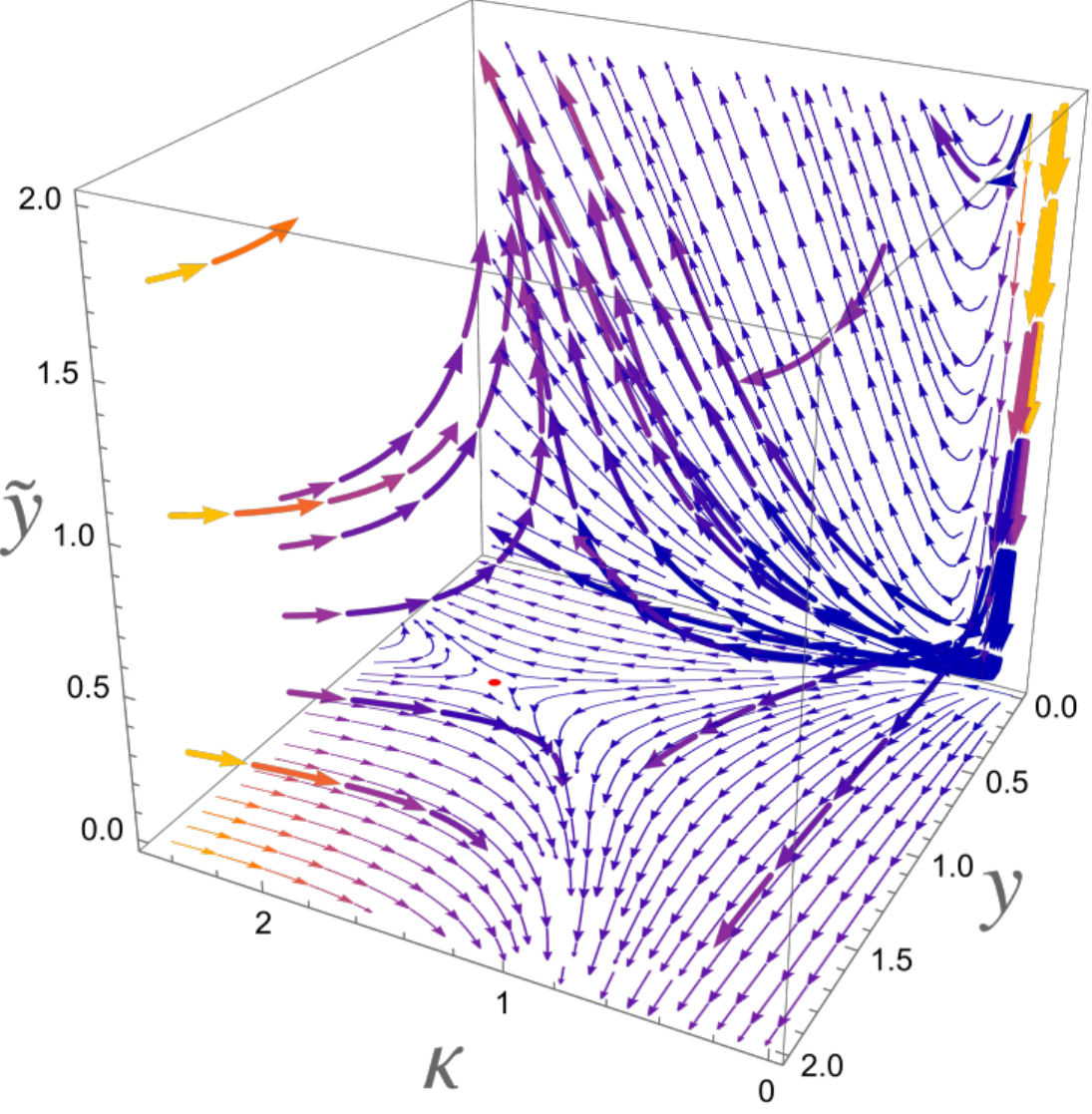}} 
	\hfill	
	\subfloat[$d=3$ $\mathcal{PT}$ broken]{\includegraphics[width=0.5\linewidth]{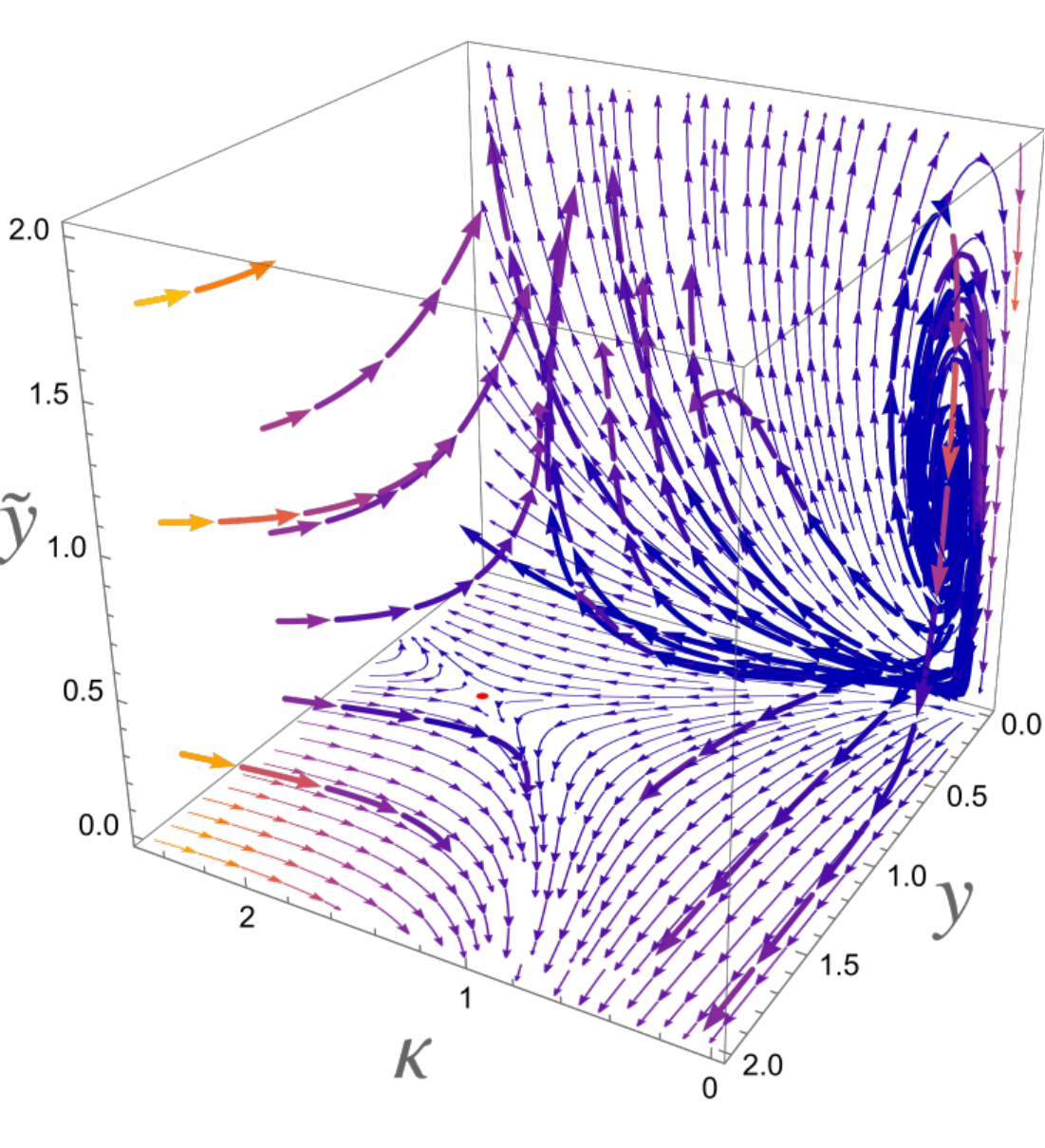}}
	\caption{RG flows for the non-Hermitian system associated to the Lagrangian of Eq.~(\ref{Eq:Leff-1}). The RG flows shown here are for $d=3$. In all panels the red dot represents the fixed point. Panels (a) and (b) represent planes featuring nontrivial fixed points arising in panels (c) and (d). In the $\mathcal{PT}$-symmetric case of panel (c) the only fixed point governs the $\rm XY$ universality class when the clock perturbation is small. In the broken $\mathcal{PT}$ symmetry case [panel (d)], both planes of panels (a) and (b) arise, with the fixed point of panel (b) being a highly unstable repulsive spiral. In this case a weakly first-order phase transition occurs.}
	\label{Fig:fig-3d}
\end{figure}

 We note that $\ln(y_r/y_i)$ is an RG invariant, so we can write $y_i= ky_r$, where $k$ is a constant. We see that for $k=1$ we obviously recover when $d=2$ the RG flow equations corresponding to a BKT behavior. In the $\mathcal{PT}$-symmetric case ($y_r>y_i$ or, equivalently, $k<1$) and $d=2$ these equations are easily shown to be equivalent to the ones obtained in Ref.~\cite{Jose_PhysRevB.16.1217}. This can be seen by deriving the RG equation for  $\widetilde{y}=\sqrt{y_r^2-y_i^2}$, which is obtained directly from Eqs.~(\ref{Eq:dyr-maintext}) and (\ref{Eq:dyi-maintext}) as, 
\begin{equation}
	\label{Eq:dy-tilde-maintext}
	\frac{d\widetilde{y}}{dl}=\left[d-\frac{4f(d)}{\pi^2\kappa}\right]\widetilde{y},
\end{equation}
along with the corresponding $d=2$ result, $d\widetilde{y}/dl=[2-4/(\pi\kappa)]\widetilde{y}$.  

If the $\mathcal{PT}$ symmetry is not broken we obtain for $d=3$ and small $\widetilde{y}$ the usual three-dimensional $\rm XY$ universality class. This critical behavior is governed by the nontrivial fixed point for $\widetilde{y}=0$ shown in panel (a) of Fig.~\ref{Fig:fig-3d}. Our RG equations yield in this case the $\epsilon$-expansion ($\epsilon=d-2$) result $\nu=1/(2\sqrt{\epsilon})+1/8+\mathcal{O}(\sqrt{\epsilon})$, which reproduces for $\epsilon=1$ one-loop value, $\nu\approx 0.625$ \cite{zinn2021quantum}. The RG flow for nonzero and large $\widetilde{y}$ exhibits a critical behavior where the four-state clock interaction becomes very large, as illustrated in panel (c) of Fig.~\ref{Fig:fig-3d}. This situation is well known to lead to two decoupled Ising models \cite{Sandvik_PhysRevLett.124.080602,Jose_PhysRevB.16.1217}.

When the $\mathcal{PT}$-symmetry is broken the situation changes considerably.  Since in this case $y_r<y_i$, Eq.~(\ref{Eq:dk-final-maintext}) becomes, 
\begin{equation}
	\label{Eq:dk-PT-broken}
	\frac{d\kappa}{dl}=-\frac{4}{\pi^2}\widetilde{y}^2-\kappa^2y^2+(d-2)\kappa,
\end{equation}
while Eqs.~(\ref{Eq:dy-maintext}) and (\ref{Eq:dy-tilde-maintext}) remain unchanged.  The first thing to note is that now we have two nontrivial fixed points, namely, $P_1=(\kappa_1,y_1,\widetilde{y}_1)=(d/f(d),\sqrt{(d-2)/\kappa_1},0)$ and  $P_2=(\kappa_2,y_2,\widetilde{y}_2)=(4f(d)/(\pi^2 d),0,(\pi/2)\sqrt{(d-2)\kappa_2})$.  The fixed point $P_1$ is the same as shown in panel (a) of Fig.~\ref{Fig:fig-3d}. The fixed point $P_2$, on the other hand, is highly unstable, corresponding to the repulsive spiral shown in panel (b) of Fig.~\ref{Fig:fig-3d}, with the full flow diagram featuring both fixed points shown in panel (d). Although the fixed point $P_1$ is the same as the one for $\widetilde{y}=0$ of panel (c) in Fig.~\ref{Fig:fig-3d}, $P_1$ {\it does not} govern the three-dimensional $\rm XY$ universality class as it is the case in panel (c). Indeed, after linearization of the RG equations in the $\mathcal{PT}$ broken regime, we obtain the eigenvalues of the linear system, $\lambda_0=d-f(d)\kappa_2$ and $\lambda_\pm=[d-2\pm i\sqrt{(d-2)(7d+2)}]/2$. The complex conjugated eigenvalues $\lambda_\pm$ are associated to the repulsive spiral fixed point $P_2$. The critical exponent $\nu$ is determined by the eigenvalue $\lambda_0$ as $\nu=1/\lambda_0$. For $d=3$ it yields $\nu=3/8$ (the $\epsilon$-expansion practically yields the same result in this case). We note that this is smaller than the mean-field value 1/2, thus indicating that a near critical behavior, or pseudocriticality, when the $\mathcal{PT}$ symmetry is broken. 

\begin{figure}
 	\subfloat[$d=2$ $\mathcal{PT}$ symmetry can be broken or not, $\widetilde{y}=0$]{\includegraphics[width=0.5\linewidth]{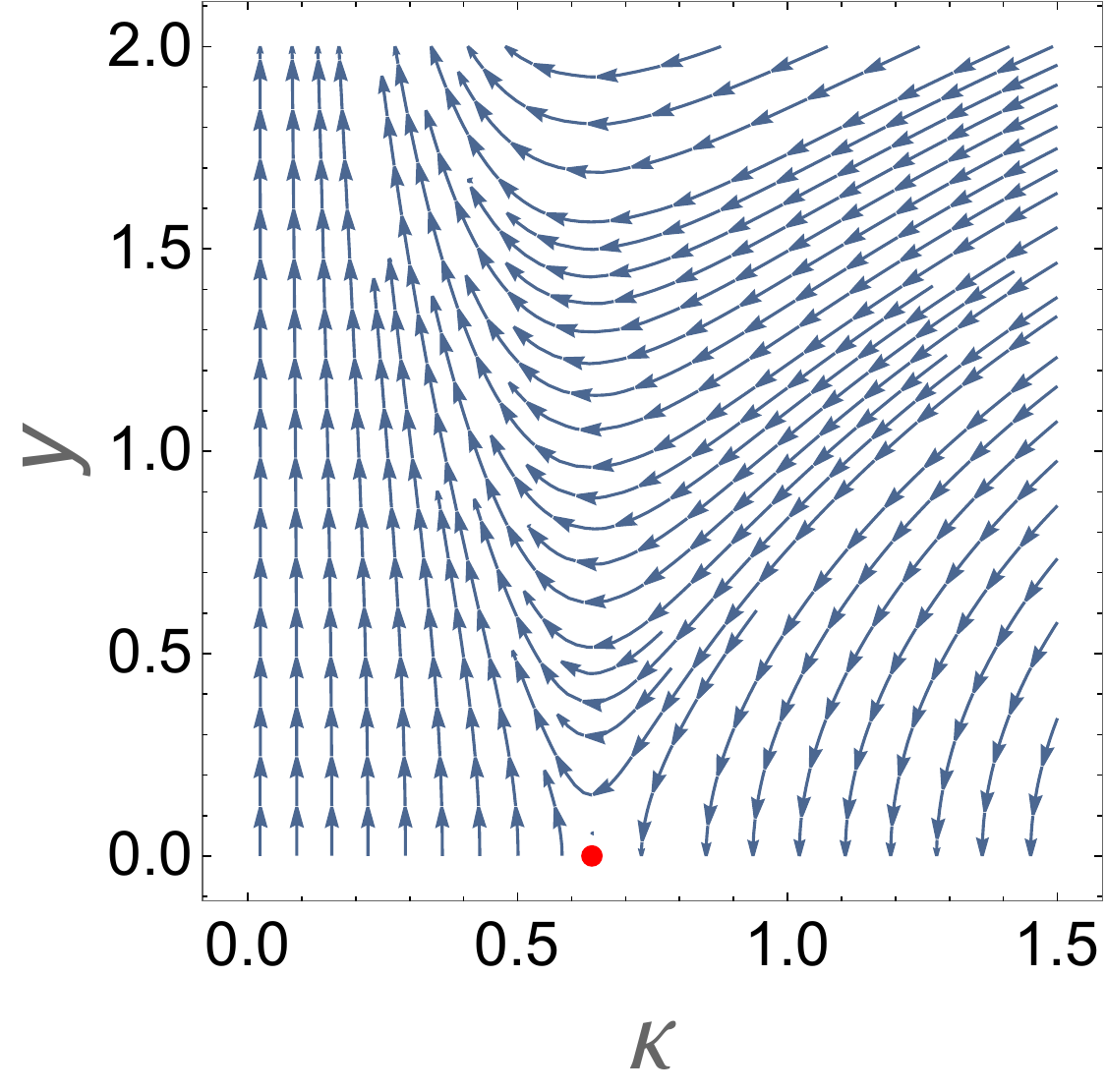}}
 	\hfill
 	\subfloat[$d=2$ $\mathcal{PT}$-symmetric, $y=0$]{\includegraphics[width=0.5\linewidth]{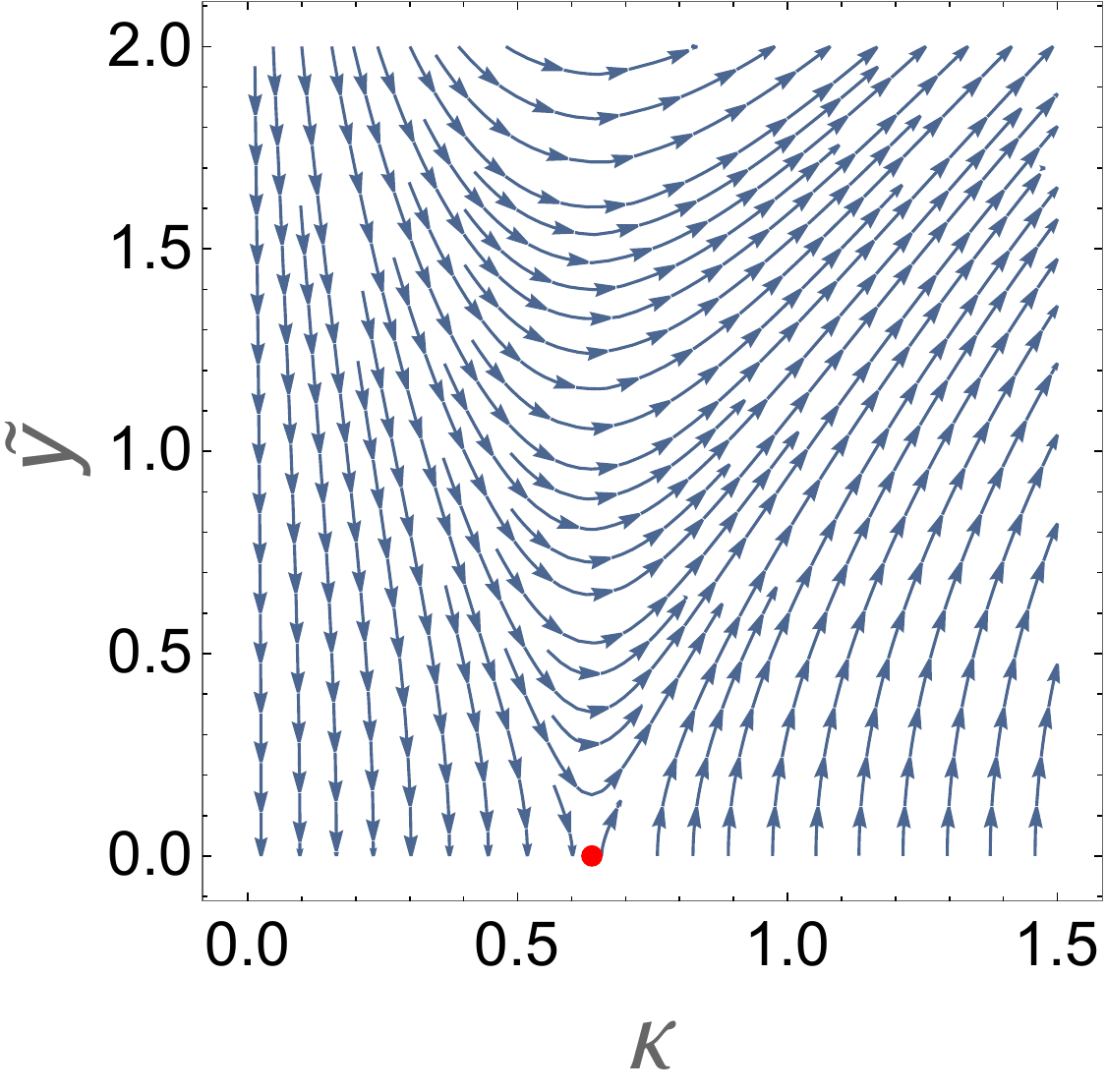}}\\
 	\subfloat[$d=2$ $\mathcal{PT}$-symmetric]{\includegraphics[width=0.5\linewidth]{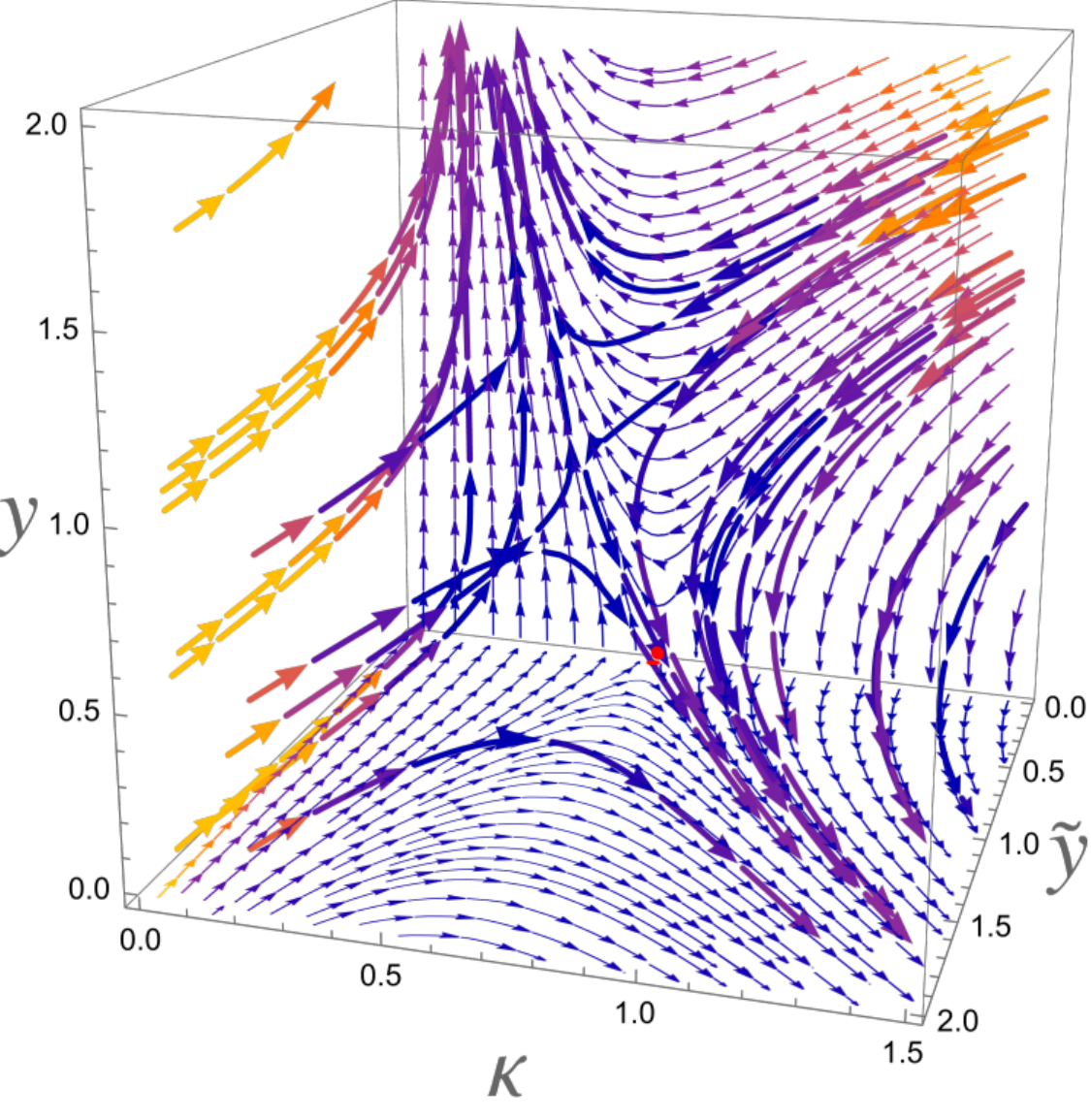}} 
 	\hfill	
 	\subfloat[$d=2$ $\mathcal{PT}$ broken]{\includegraphics[width=0.5\linewidth]{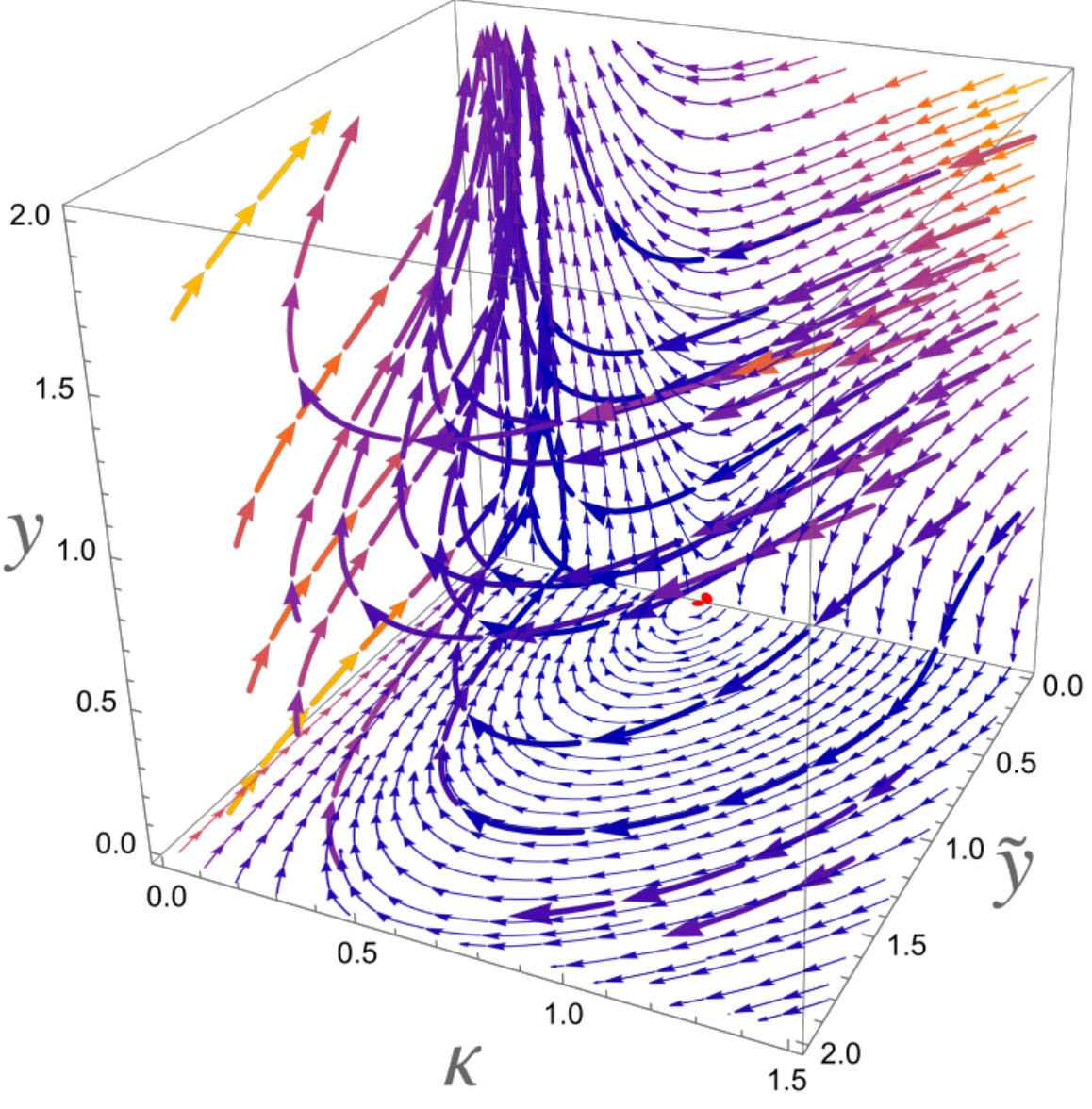}}
 	\hfill
 	\subfloat[RG-invariant surfaces $\mathcal{PT}$ broken]{\includegraphics[width=0.5\linewidth]{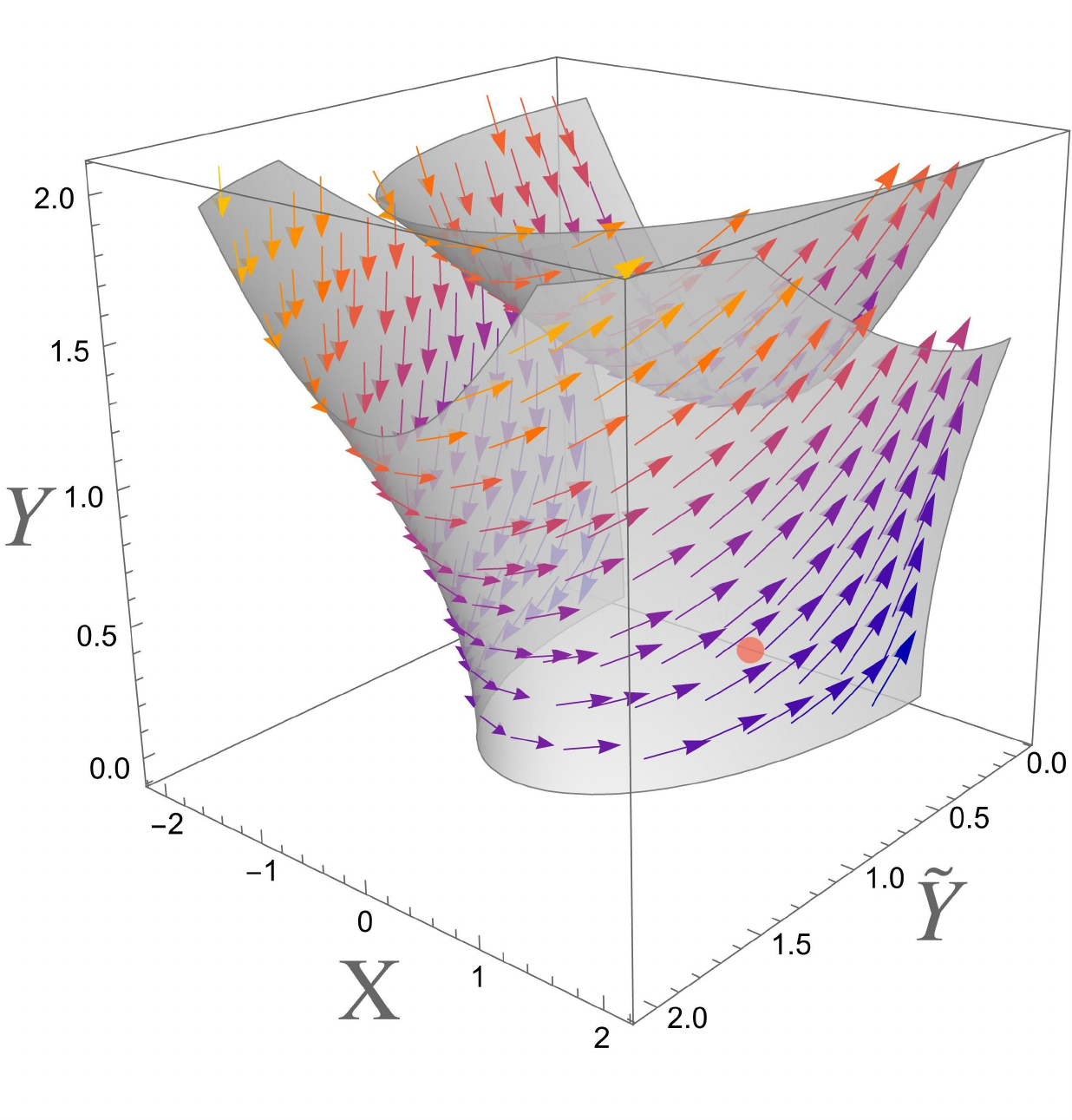}}
 	\hfill
 	\subfloat[Conical separatrix $\mathcal{PT}$ broken]{\includegraphics[width=0.5\linewidth]{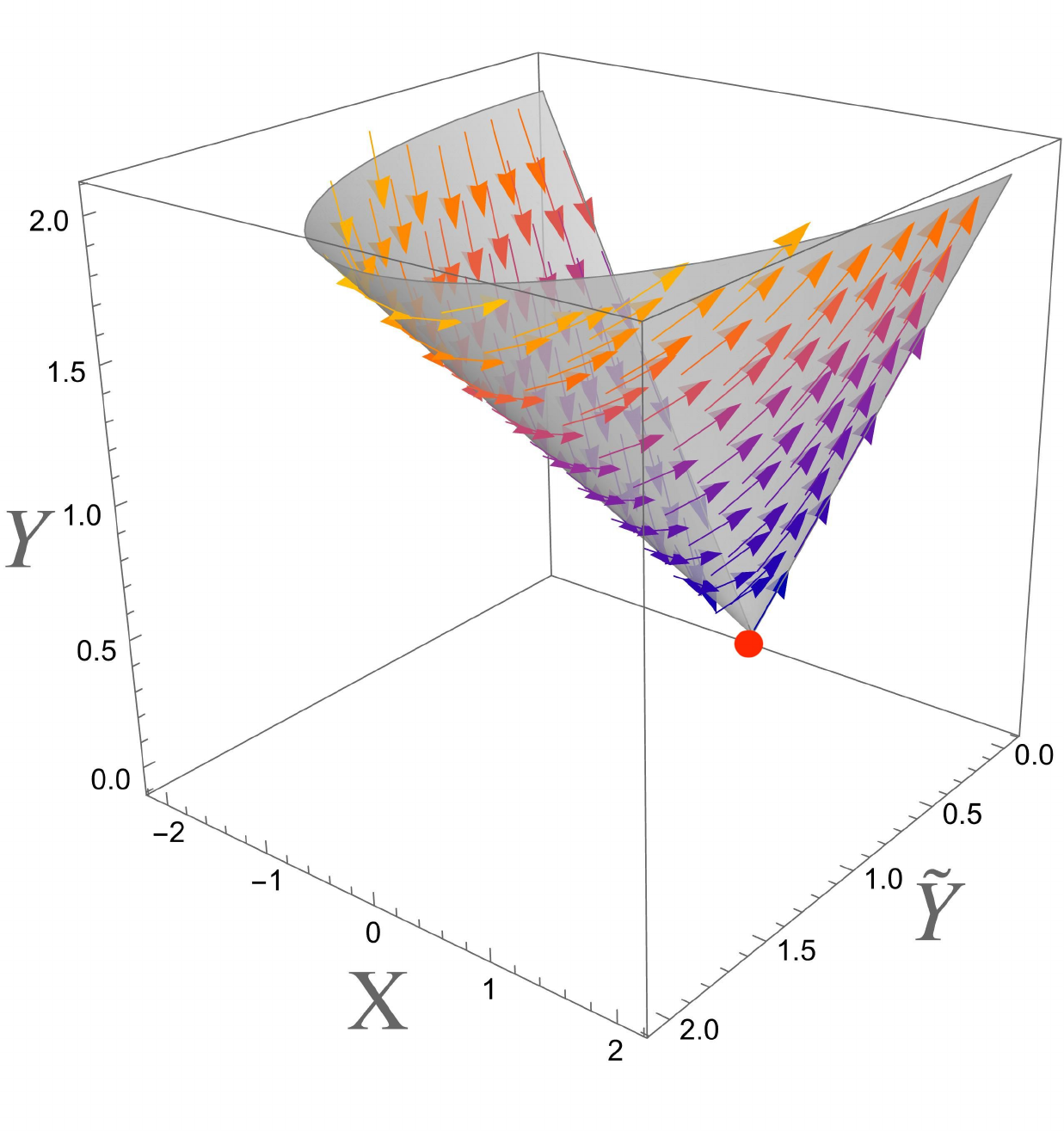}}
 	\caption{RG flows for the non-Hermitian system associated to the Lagrangian of Eq.~(\ref{Eq:Leff-1}). The RG flows shown here are for $d=2$. In all panels the red dot represents the fixed point $\kappa_*=2/\pi$ and $\widetilde{y}=y=0$. Panels (a) and (b) show separately the BKT planes appearing in panel (c) for the $\mathcal{PT}$-symmetric case. Note that these planes are dual to each other in the sense that the fixed line is for $\kappa\geq \kappa_*$ in panel (a) and for $\kappa\leq \kappa_*$ in panel (b). When the $\mathcal{PT}$ symmetry is broken only the BKT plane of panel (a) survives, causing the BKT transition seems to dominate the critical behavior, as shown in panel (d). Panel (e) shows the RG-invariant surfaces. In panel (f)  the conical separatrix with apex at the fixed point is shown. In both panels (e) and (f) we have defined the variables, $X=2-\pi\kappa$, $Y=2y/\sqrt{\pi}$, and $\widetilde{Y}=2\widetilde{y}/\sqrt{\pi}$.} 
 	\label{Fig:fig-2d}
 \end{figure}

When $d\to 2$ the fixed points $P_1$ and $P_2$ collide, representing a typical characteristic of walking behavior \cite{Rychkov-Walking}. In Fig.~\ref{Fig:fig-2d} we compare the scaling behaviors of the $\mathcal{PT}$-symmetric with the $\mathcal{PT}$ broken case. In the $\mathcal{PT}$-symmetric case both $dy/dl$ and $d\widetilde{y}/dl$ vanish for $\kappa_*=2/\pi$, leading to three lines of fixed points, as $\widetilde{y}^2_*=y^2_*$ provides one more fixed line in addition to the two BKT ones determined by the vanishing of both $y$ and $\widetilde{y}$. In panels (a) and (b) of Fig.~\ref{Fig:fig-2d} we illustrate the dual RG BKT flows for the situations where $\widetilde{y}=0$ and $y=0$, respectively. Overall, however, the phase transition is not a BKT one in this case: the correlation length exhibits a power-law with a continuously varying critical exponent $\nu(y_*)\sim 1/y_*$ \cite{Jose_PhysRevB.16.1217}. Panel (c) of Fig.~\ref{Fig:fig-2d} shows a view of the full RG flow diagram. {\color{black} The behavior of the correlation function, 
\begin{equation}
	\label{Eq:Correlfunc}
	G(x,x')=\langle e^{i[\theta(x)-\theta(x')]}\rangle,
\end{equation}
is well known in this case \cite{Jose_PhysRevB.16.1217,KADANOFF197939}. At the order we are calculating the anomalous dimension $\eta$ of the critical correlation function does not differ from the one found in the BKT transition, namely, $\eta=1/4$, even though the transition is not a BKT one, since we are considering a four-state clock model \cite{Jose_PhysRevB.16.1217}. Away from the critical point the scaling behavior is governed by the correlation length, which as mentioned above, features a continuously varying critical exponent $\nu(y_*)\sim 1/y_*$. While no local order parameter can be defined for $\rm XY$ model leading to a BKT transition, there is no such a Mermin-Wagner constraint in the case of the four-state clock model in view of the $Z_4$ symmetry. Therefore, the order parameter behaves as $\langle e^{i\theta}\rangle\sim (\kappa_*-\kappa)^{\nu\eta/2}$. Thus, we find that the critical exponent $\beta=\nu\eta/2$ for the order parameter is also continuously varying.}

In the $\mathcal{PT}$ broken regime a BKT flow is still present for $\widetilde{y}=0$, but the BKT flow for $y=0$ disappears, as it is illustrated in panel (d) of Fig.~\ref{Fig:fig-2d}. The resulting RG flow is better understood by introducing the variables  $X=2-\pi\kappa$, $Y=2y/\sqrt{\pi}$, and $\widetilde{Y}=2\widetilde{y}/\sqrt{\pi}$, in which case we can rewrite the RG equations in the form, $dX/dl\approx \widetilde{Y}^2+Y^2$, $dY/dl=XY$, and $d\widetilde{Y}/dl\approx-X\widetilde{Y}$. We obtain that the family of hyperboloids, $X^2-Y^2+\widetilde{Y}^2=c^2={\rm const}$, are RG-invariants, see Fig.~\ref{Fig:fig-2d}-(e). At the critical point the constant $c$ vanishes, thus defining the cone with apex at the fixed point shown in Fig.~\ref{Fig:fig-2d}-(f) as a separatrix. Using the hyperboloid RG invariant one can proceed to approximately solve the RG flow equations to obtain the correlation length. For details, see Appendix B. From the solution, $X(l)=X(0)\cos(\sqrt{2}cl)$, the correlation length $\xi$ is obtained by demanding that $X(l_*)\sim\mathcal{O}(1)$ for some scale $l_*$ such that $\ln(\xi/a)=l_*$. This leads to a BKT scaling, $\xi^{-1}\sim \exp(-{\rm const}/\sqrt{K-K_c})$, where $K_c=\kappa_*$.  The critical behavior is thus essentially governed by the BKT flow diagram of Fig.~\ref{Fig:fig-2d}-(b). Strictly speaking the phase transition is not exactly a BKT one though. The flow lines do not stop at the fixed line that exists for $\widetilde{Y}=0$ when the latter is nonzero, no matter how small. They follow a parabolic trajectory around the conical separatrix shown in Fig.~\ref{Fig:fig-2d}-(f). In other words, it is a BKT-like scaling occurring by means of a running around the fixed point, with no fixed lines present.  In view of these results the $\mathcal{PT}$ broken regime 
evidences a scaling behavior known as walking or pseudocritical, which was studied in detail by Gorbenko {\it et al.} \cite{Rychkov-Walking} (see also Refs.~\cite{Herbut_PhysRevD.94.025036,Nahum_PhysRevX.5.041048,Scherer_PhysRevB.100.134507,Weber-Vojta_PhysRevLett.130.186701}, a concept closely related to 'conformality loss'
\cite{Miransky_PhysRevD.55.5051,Son_PhysRevD.80.125005,Nogueira_2013,Nogueira_PhysRevD.100.085005,Ma-He_PhysRevB.99.195130}).  

{\color{black} In the $\mathcal{PT}$ broken regime which follows from the fixed point collision as $d\to 2$, the correlation function exhibits a critical behavior similar to the one discussed above, with the same anomalous dimension. However, the walking BKT-like scaling implies that the order parameter always vanishes, similarly to the usual BKT transition. } 

\section{Conclusions and outlook} 
We have determined the phase structure and critical behavior of a non-Hermitian $\rm XY$ model with a four-state clock perturbation both in 1+1 and 2+1 dimensions. Our RG analysis gives information about both $\mathcal{PT}$-symmetric and $\mathcal{PT}$-broken regimes, where the former phases are associated with real energy eigenvalues giving rise to a unitary time-evolution and the latter not.  In view of the exact equivalence of the $\mathcal{PT}$-symmetric Lagrangian to a Hermitian clock model, our RG analysis in this case essentially corroborates existing numerical works for the Hermitian clock model \cite{Hove-Sudbo_PhysRevE.68.046107,Sandvik_PhysRevLett.124.080602,Sandvik_PhysRevB.103.054418}.  On the other hand, our RG analysis also gives access to the $\mathcal{PT}$-broken regime and we show that its phase structure and critical behavior are very different from its $\mathcal{PT}$-symmetric counterpart, as discussed in the main text. The main point we make here is that the $\mathcal{PT}$ broken regime exhibits walking (or pseudocritical) behavior in the sense defined in Ref.~\cite{Rychkov-Walking}. 

This analysis paves the way to consider the role of dangerously irrelevant operators \cite{AMIT1982207} in non-Hermitian systems, which are pertinent for non-Hermitian deconfined quantum critical points, where in a number of cases can be qualified as pseudocritical \cite{Rychkov-Walking,Nogueira_2013,Nahum_PhysRevX.5.041048,Scherer_PhysRevB.100.134507}. 

From an experimental point of view, we expect that the effective theories investigated here may for instance be realized using Josephson junction arrays with asymmetric current biases \cite{Marcus}.
Other interesting experimental possibilities arise when certain non-Hermitian conductance matrices are interpreted as non-Hermitian Hamiltonians, as recently discussed in Refs.~\cite{Kiril,koenye2023nonhermitian}.      

\begin{acknowledgments}
We thank Slava Rychkov for useful comments and discussions on the concept of walking. 
	We acknowledge financial support by the Deutsche Forschungsgemeinschaft (DFG, German Research Foundation), through SFB 1143 project A5 and the W{\"u}rzburg-Dresden Cluster of Excellence on Complexity and Topology in Quantum Matter-ct.qmat (EXC 2147, Project Id No. 390858490). 
\end{acknowledgments}

\appendix
\section{Derivation of the RG equations}

Given the renormalized counterparts $\kappa$ and $y$ of $K$ and $z$, respectively, we obtain the usual RG equations associated to the BKT phase transition \cite{Nelson-Kosterlitz1977,Jose_PhysRevB.16.1217},
\begin{equation}
	\label{Eq:RG-k}
	\frac{d\kappa^{-1}}{dl}=y^2,
\end{equation}
\begin{equation}
	\label{Eq:RG-y}
	\frac{dy}{dl}=(2-\pi\kappa)y. 
\end{equation}

Switching on the $n=4$ clock model perturbation modifies these RG equations considerably \cite{Jose_PhysRevB.16.1217}. In order to obtain the result, we simply disregard the periodicity of $\phi$ in Eq.~(\ref{Eq:Leff-2}), which makes it equivalent to an ordinary sine-Gordon theory, which is known to be exactly equivalent to a Coulomb gas in $d$ dimensions \cite{frohlich1981statistical}. Note that in this case we are not specializing to $d=2$. The reason for this will become clear below. Importantly, we have to adapt the derivation as given in Ref.~\cite{Nogueira-Kleinert_PhysRevB.77.045107} such that it allows for a non-Hermitian sine-Gordon interaction. Crucial in this regard is the role played by the spatially dependent dielectric constant, 
\begin{equation}
	\varepsilon(r)=1+S_d\chi(r),
\end{equation}
where $S_d=2\pi^{d/2}/\Gamma(d/2)$ is the surface area of the unit sphere in $d$ dimensions and the electric susceptibility of the Coulomb gas is given by, 
\begin{equation}
	\chi(r)=S_d\int_a^r dss^{d-1}\alpha(s)n(s).
\end{equation} 
In the above expression $\alpha(r)$ represents the polarizability and $n(r)$ is the local average number of dipoles. It is in this quantity that the couplings $z_r$ and $z_i$ appearing in Eq.~(\ref{Eq:Leff-1}) of the main text will enter. Approximately (i.e., for small dipoles) we have, 
\begin{equation}
	\label{Eq:n}
	n(r)\approx (z_r^2-z_i^2)e^{-U(r)},
\end{equation}
where $U(r)$ is the renormalized electric potential. From the self-consistent analysis of $U(r)$ the RG equations (\ref{Eq:dyr-maintext}) and (\ref{Eq:dyi-maintext}) in the main text are obtained for the dimensionless couplings $y_r$ and $y_i$ associated to $z_r$ and $z_i$. Note that for broken $\mathcal{PT}$ symmetry where $z_r<z_i$ the density (\ref{Eq:n}) becomes negative (in this sense $n(r)$ is best interpreted as a density imbalance), and the dielectric constant $\varepsilon(r)<1$, leading to an anti-screening in the Coulomb gas. 

Defining $\kappa=4/(\pi^2\widetilde{\kappa})$ and $\widetilde{y}$ as the dimensionless renormalized counterpart of $\widetilde{z}$, we obtain using the results of Refs.~\cite{Kosterlitz_1977,KLEINERT2003361,Nogueira-Kleinert_PhysRevB.77.045107} the leading order RG equations in $d$ dimensions in the form, 
\begin{equation}
	\label{Eq:RG-K-tilde}
	\frac{d\widetilde{\kappa}^{-1}}{dl}=\widetilde{y}^2+(d-2)\widetilde{\kappa}^{-1},
\end{equation}
\begin{equation}
	\label{Eq:dy-tilde}
	\frac{d\widetilde{y}}{dl}=\left[d-f(d)\widetilde{\kappa}\right]\widetilde{y}, 
\end{equation}
where $f(d)=(d-2)\Gamma(d/2-1)/(2\pi^{d/2-2})$ and $l=\ln(r/a)$ is the RG logarithmic scale. 

In order to obtain the full set of RG equations for $d>2$ we have to surmount the technical difficulty imposed by the fact that in this case we are not dealing with an ensemble of point vortices any longer. In fact, for $d=3$ the duality leads to an Abelian Higgs theory describing an ensemble of vortex loops rather than a sine-Gordon theory \cite{Peskin1978,Dasgupta-Halperin_PhysRevLett.47.1556,kleinert1989gauge}. However, it has been argued in the past \cite{Nelson-Fisher_PhysRevB.16.4945,Cardy-Hamber_PhysRevLett.45.499} that near $d=2$ the RG equations for $d>2$ simply amount to a modification of Eq.~(\ref{Eq:RG-k}) in which a term $-(d-2)\kappa^{-1}$ is added to its right-hand side. Crucially, we note that this term has a sign opposite to the similar term appearing in Eq.~(\ref{Eq:RG-K-tilde}) for the Coulomb gas case. The latter leads to the well known result that the Coulomb gas for $d>2$ does not undergo a phase transition, reflecting the absence of a fixed point and the runaway flow \cite{Kosterlitz_1977,Nogueira-Kleinert_PhysRevB.77.045107}. 

Doing an $\epsilon$-expansion ($\epsilon=d-2$) in the way just mentioned, which follows the prescription in Refs.~\cite{Nelson-Fisher_PhysRevB.16.4945,Cardy-Hamber_PhysRevLett.45.499} has the drawback that it yields a correlation length critical exponent $\nu=1/(2\sqrt{\epsilon})[1+\mathcal{O}(\sqrt{\epsilon})]$, which does not deviate from the mean-field theory value $\nu=1/2$ when $\epsilon=1$. However, this can be improved in the following way. We can argue that the RG equation in $d$ dimensions for $y$ has the same form as the one for $\widetilde{y}$ given in Eq.~(\ref{Eq:dy-tilde}), i.e., we do not perform the $\epsilon$-expansion in the RG equation. This would lead to the following RG equations in absence of the symmetry breaking field, 
\begin{equation}
	\label{Eq:RG-K-d}
	\frac{d\kappa^{-1}}{dl}=y^2-(d-2)\kappa^{-1},
\end{equation}
\begin{equation}
	\label{Eq:dy-d}
	\frac{dy}{dl}=\left[d-f(d)\kappa\right]y. 
\end{equation}
Linearizing the above equations around the fixed point with coordinates $\kappa_*=d/f(d)$ and $y_*=\sqrt{(d-2)/\kappa_*}$, we find the eigenvalues $a_\pm=[2-d\pm\sqrt{(d-2)(9d-2)}]/2$, leading to $1/\nu=a_+$. While setting $d=3$ directly in this expression yields exactly $\nu=1/2$, we have to remember that the procedure assumes that near $d=2$ we can still use point vortices as a leading order approximation to vortex loops. Hence, our expression for $\nu$ should also be considered within the framework of an $\epsilon$-expansion. As a matter of fact, we obtain, $\nu=1/(2\sqrt{\epsilon})+1/8+\mathcal{O}(\sqrt{\epsilon})$, which in turns yields $\nu\approx 0.625$ when we set $\epsilon=1$ at the end. This result precisely agrees with the one-loop RG result obtained from a Landau-Ginzburg-Wilson analysis of a $|\psi|^4$ theory. There are other approaches to improve the strategy followed by Nelson and Fisher \cite{Nelson-Fisher_PhysRevB.16.4945}, like for example dealing more directly with the scaling behavior of vortex loops \cite{Shenoy_PhysRevB.40.5056}, which ultimately leads to the more accurate result, $\nu\approx 2/3$ \cite{zinn2021quantum}. However, our approach has the virtue of being technically simpler, as it relies more on the original idea of Refs.~\cite{Nelson-Fisher_PhysRevB.16.4945,Cardy-Hamber_PhysRevLett.45.499}. 

After the above preliminary discussion, we are now finally in position to write down the RG equations for the Lagrangian of Eq.~(\ref{Eq:Leff-2}), which simply amounts to combining the results of Eqs.~(\ref{Eq:RG-K-tilde}) and (\ref{Eq:dy-tilde}) with those of Eqs.~(\ref{Eq:RG-K-d}) and (\ref{Eq:dy-d}) to obtain,
\begin{equation}
	\label{Eq:dk-final}
	\frac{d\kappa}{dl}=\frac{4}{\pi^2}\widetilde{y}^2-\kappa^2y^2+(d-2)\kappa,
\end{equation}
\begin{equation}
	\label{Eq:dy-tilde-final}
	\frac{d\widetilde{y}}{dl}=\left[d-\frac{4f(d)}{\pi^2\kappa}\right]\widetilde{y},
\end{equation}          
while the flow equation for $y$ is given by Eq.~(\ref{Eq:dy-d}). 

\begin{widetext}
	
\section{Aproximate analysis of the BKT-like scaling in the $\mathcal{PT}$ broken regime}

RG equations for $\mathcal{PT}$ broken region in $1+1$ dimensions are given by,
\begin{equation}
	\label{Eq:dk-PT-broken-d=2}
	\frac{d\kappa}{dl}=-\frac{4}{\pi^2}\widetilde{y}^2-\kappa^2y^2, \qquad \frac{dy}{dl}=\left[2-\pi\kappa\right]y, \qquad \frac{d\widetilde{y}}{dl}=\left[2-\frac{4}{\pi\kappa}\right]\widetilde{y}.
\end{equation}
Let's introduce the new variables,
\begin{equation}
	X=2-\pi\kappa, \qquad Y=\frac{2}{\sqrt{\pi}}y, \qquad \widetilde{Y}=\frac{2}{\sqrt{\pi}}\widetilde{y}.
\end{equation}
The RG equations become
\begin{equation}
	\label{Eq:dX-PT-broken}
	\frac{dX}{dl}=\widetilde{Y}^2+\left(1-\frac{X}{2}\right)^2Y^2\approx\widetilde{Y}^2+Y^2, \qquad \frac{dY}{dl}=XY, \qquad \frac{d\widetilde{Y}}{dl}=2\left(1-\frac{1}{1-\frac{X}{2}}\right)\widetilde{Y}\approx-X\widetilde{Y}.
\end{equation}
Thus, the family of hyperboloid
\begin{equation}
	X^2-Y^2+\widetilde{Y}^2=c^2=\text{const}
\end{equation}
are RG invariants. For $X^2+\widetilde{Y}^2<Y^2$, we parametrize the hyperboloid of two sheets by
\begin{equation}
	X=c\sinh{u}\cos{v}, \qquad Y=c\cosh{u}, \qquad \widetilde{Y}=c\sinh{u}\sin{v}, \qquad u\in\mathbb{R}^2, \quad v\in\left[0,\pi\right).
\end{equation}
It is convenient to define new variable
\begin{equation}
	X(l)\approx X_0+\widetilde{X}(l).
\end{equation}
and rewrite the Eqs.~\eqref{Eq:dX-PT-broken} in terms of $X_0$ and $\widetilde{X}$
\begin{equation}
	\label{Eq:dTildeX-PT-broken}
	\frac{d\widetilde{X}}{dl}\approx\widetilde{Y}^2+Y^2, \qquad \frac{dY}{dl}\approx X_0 Y, \qquad \frac{d\widetilde{Y}}{dl}\approx-X_0\widetilde{Y}.
\end{equation}
In this approximation, we can obtain a solution in closed form
\begin{equation}
	u(l)=\text{arccosh}\left(e^{X_0 l}\right), \qquad v(l)=\arcsin\left(\frac{e^{-2X_0 l}}{\sqrt{e^{-2X_0 l}-1}}\right).
\end{equation}
In the large-distance limit, $l$ goes to infinity. Therefore, if $X_0$ is positive, then $u(l)$ increases and $v(l)$ goes to $0$. As a result $X(l)$ and $Y(l)$ also increas , until the RG trajectory leaves the perturbative regime and $\widetilde{Y}(l)$ goes to $0$. By contrast, if $X_0<0$ then $u(l)$ goes to $0$ and $v(l)$ increases. In this case, our approximation does not work, since $v$ is limited by the interval $\left[0,\pi\right)$.

For $X^2+\widetilde{Y}^2>Y^2$, we parametrize the hyperboloid of one sheet by
\begin{equation}
	X=c\cosh{u}\cos{v}, \qquad Y=c\sinh{u}, \qquad \widetilde{Y}=c\cosh{u}\sin{v}, \qquad u\in\mathbb{R}^2, \quad v\in\left[0,\pi\right).
\end{equation}
The solution of Eqs.~\eqref{Eq:dTildeX-PT-broken} is then
\begin{equation}
	u(l)=\text{arcsinh}\left(e^{X_0 l}\right), \qquad v(l)=\arcsin\left(\frac{e^{-2X_0 l}}{\sqrt{X_0(e^{-2X_0 l}+1)}}\right).
\end{equation}
Therefore, if $X_0>0$, then we have exactly the same behavior as in the previous case.

Since the critical point is characterized by $X=0$, $Y=0$ and $\widetilde{Y}=0$, the constant $c^2$ has to vanish there. Thus, $c^2$ controls the distance to the critical point, which we can capture in terms of the parameter $K$, according to
\begin{equation}
	c^2=b\left(K-K_c\right),
\end{equation}
where $b$ is an unimportant prefactor. Using Eqs.~\eqref{Eq:dX-PT-broken}, we can write
\begin{equation}
	\label{Eq:d^2X-PT-broken}
	\frac{d^2X}{dl^2}=2\widetilde{Y}\frac{d\widetilde{Y}}{dl}+2Y\frac{dY}{dl}=-2X(-Y^2+\widetilde{Y}^2)=-2X(c^2-X^2)\approx-2c^2X,
\end{equation}
which can be satisfied by the next ansatz
\begin{equation}
	X(l)=X(0)\cos(\sqrt{2}cl).
\end{equation}
We evaluate the left-hand side when $X(l)$ is of order 1. This happens at a scale $l=l_*$ which is related to the correlation length $\xi=a e^{l_*}$. Equation
\begin{equation}
	X(0)\cos(\sqrt{2}cl_*)=1
\end{equation}
has a solution
\begin{equation}
	l_*=\frac{1}{\sqrt{2b(K-K_c)}}\arccos\left[1/X(0)\right], \qquad |X(0)|\geqslant 1
\end{equation}
and finally
\begin{equation}
	\xi^{-1}=a^{-1}\exp\left(-\frac{1}{\sqrt{2b(K-K_c)}}\arccos\left[1/X(0)\right]\right).
\end{equation}

\end{widetext}

\bibliography{nh-qcp}

\begin{thebibliography}{65}%
\makeatletter
\providecommand \@ifxundefined [1]{%
 \@ifx{#1\undefined}
}%
\providecommand \@ifnum [1]{%
 \ifnum #1\expandafter \@firstoftwo
 \else \expandafter \@secondoftwo
 \fi
}%
\providecommand \@ifx [1]{%
 \ifx #1\expandafter \@firstoftwo
 \else \expandafter \@secondoftwo
 \fi
}%
\providecommand \natexlab [1]{#1}%
\providecommand \enquote  [1]{``#1''}%
\providecommand \bibnamefont  [1]{#1}%
\providecommand \bibfnamefont [1]{#1}%
\providecommand \citenamefont [1]{#1}%
\providecommand \href@noop [0]{\@secondoftwo}%
\providecommand \href [0]{\begingroup \@sanitize@url \@href}%
\providecommand \@href[1]{\@@startlink{#1}\@@href}%
\providecommand \@@href[1]{\endgroup#1\@@endlink}%
\providecommand \@sanitize@url [0]{\catcode `\\12\catcode `\$12\catcode
  `\&12\catcode `\#12\catcode `\^12\catcode `\_12\catcode `\%12\relax}%
\providecommand \@@startlink[1]{}%
\providecommand \@@endlink[0]{}%
\providecommand \url  [0]{\begingroup\@sanitize@url \@url }%
\providecommand \@url [1]{\endgroup\@href {#1}{\urlprefix }}%
\providecommand \urlprefix  [0]{URL }%
\providecommand \Eprint [0]{\href }%
\providecommand \doibase [0]{https://doi.org/}%
\providecommand \selectlanguage [0]{\@gobble}%
\providecommand \bibinfo  [0]{\@secondoftwo}%
\providecommand \bibfield  [0]{\@secondoftwo}%
\providecommand \translation [1]{[#1]}%
\providecommand \BibitemOpen [0]{}%
\providecommand \bibitemStop [0]{}%
\providecommand \bibitemNoStop [0]{.\EOS\space}%
\providecommand \EOS [0]{\spacefactor3000\relax}%
\providecommand \BibitemShut  [1]{\csname bibitem#1\endcsname}%
\let\auto@bib@innerbib\@empty
\bibitem [{\citenamefont {Bender}\ and\ \citenamefont
  {Boettcher}(1998)}]{Bender_PhysRevLett.80.5243}%
  \BibitemOpen
  \bibfield  {author} {\bibinfo {author} {\bibfnamefont {C.~M.}\ \bibnamefont
  {Bender}}\ and\ \bibinfo {author} {\bibfnamefont {S.}~\bibnamefont
  {Boettcher}},\ }\bibfield  {title} {\bibinfo {title} {Real spectra in
  non-hermitian hamiltonians having $\mathcal{PT}$ symmetry},\ }\href
  {https://doi.org/10.1103/PhysRevLett.80.5243} {\bibfield  {journal} {\bibinfo
   {journal} {Phys. Rev. Lett.}\ }\textbf {\bibinfo {volume} {80}},\ \bibinfo
  {pages} {5243} (\bibinfo {year} {1998})}\BibitemShut {NoStop}%
\bibitem [{\citenamefont {Bender}(2005)}]{Bender-Introduction}%
  \BibitemOpen
  \bibfield  {author} {\bibinfo {author} {\bibfnamefont {C.~M.}\ \bibnamefont
  {Bender}},\ }\bibfield  {title} {\bibinfo {title} {Introduction to
  $\mathcal{PT}$-symmetric quantum theory},\ }\href
  {https://doi.org/10.1080/00107500072632} {\bibfield  {journal} {\bibinfo
  {journal} {Contemporary Physics}\ }\textbf {\bibinfo {volume} {46}},\
  \bibinfo {pages} {277} (\bibinfo {year} {2005})},\ \Eprint
  {https://arxiv.org/abs/https://doi.org/10.1080/00107500072632}
  {https://doi.org/10.1080/00107500072632} \BibitemShut {NoStop}%
\bibitem [{\citenamefont {Ashida}\ \emph {et~al.}(2020)\citenamefont {Ashida},
  \citenamefont {Gong},\ and\ \citenamefont {Ueda}}]{Ashida-Review}%
  \BibitemOpen
  \bibfield  {author} {\bibinfo {author} {\bibfnamefont {Y.}~\bibnamefont
  {Ashida}}, \bibinfo {author} {\bibfnamefont {Z.}~\bibnamefont {Gong}},\ and\
  \bibinfo {author} {\bibfnamefont {M.}~\bibnamefont {Ueda}},\ }\bibfield
  {title} {\bibinfo {title} {Non-hermitian physics},\ }\href
  {https://doi.org/10.1080/00018732.2021.1876991} {\bibfield  {journal}
  {\bibinfo  {journal} {Advances in Physics}\ }\textbf {\bibinfo {volume}
  {69}},\ \bibinfo {pages} {249} (\bibinfo {year} {2020})},\ \Eprint
  {https://arxiv.org/abs/https://doi.org/10.1080/00018732.2021.1876991}
  {https://doi.org/10.1080/00018732.2021.1876991} \BibitemShut {NoStop}%
\bibitem [{\citenamefont {Stegmaier}\ \emph {et~al.}(2021)\citenamefont
  {Stegmaier}, \citenamefont {Imhof}, \citenamefont {Helbig}, \citenamefont
  {Hofmann}, \citenamefont {Lee}, \citenamefont {Kremer}, \citenamefont
  {Fritzsche}, \citenamefont {Feichtner}, \citenamefont {Klembt}, \citenamefont
  {H\"ofling}, \citenamefont {Boettcher}, \citenamefont {Fulga}, \citenamefont
  {Ma}, \citenamefont {Schmidt}, \citenamefont {Greiter}, \citenamefont
  {Kiessling}, \citenamefont {Szameit},\ and\ \citenamefont
  {Thomale}}]{Thomale_PhysRevLett.126.215302}%
  \BibitemOpen
  \bibfield  {author} {\bibinfo {author} {\bibfnamefont {A.}~\bibnamefont
  {Stegmaier}}, \bibinfo {author} {\bibfnamefont {S.}~\bibnamefont {Imhof}},
  \bibinfo {author} {\bibfnamefont {T.}~\bibnamefont {Helbig}}, \bibinfo
  {author} {\bibfnamefont {T.}~\bibnamefont {Hofmann}}, \bibinfo {author}
  {\bibfnamefont {C.~H.}\ \bibnamefont {Lee}}, \bibinfo {author} {\bibfnamefont
  {M.}~\bibnamefont {Kremer}}, \bibinfo {author} {\bibfnamefont
  {A.}~\bibnamefont {Fritzsche}}, \bibinfo {author} {\bibfnamefont
  {T.}~\bibnamefont {Feichtner}}, \bibinfo {author} {\bibfnamefont
  {S.}~\bibnamefont {Klembt}}, \bibinfo {author} {\bibfnamefont
  {S.}~\bibnamefont {H\"ofling}}, \bibinfo {author} {\bibfnamefont
  {I.}~\bibnamefont {Boettcher}}, \bibinfo {author} {\bibfnamefont {I.~C.}\
  \bibnamefont {Fulga}}, \bibinfo {author} {\bibfnamefont {L.}~\bibnamefont
  {Ma}}, \bibinfo {author} {\bibfnamefont {O.~G.}\ \bibnamefont {Schmidt}},
  \bibinfo {author} {\bibfnamefont {M.}~\bibnamefont {Greiter}}, \bibinfo
  {author} {\bibfnamefont {T.}~\bibnamefont {Kiessling}}, \bibinfo {author}
  {\bibfnamefont {A.}~\bibnamefont {Szameit}},\ and\ \bibinfo {author}
  {\bibfnamefont {R.}~\bibnamefont {Thomale}},\ }\bibfield  {title} {\bibinfo
  {title} {Topological defect engineering and $\mathcal{P}\mathcal{T}$ symmetry
  in non-hermitian electrical circuits},\ }\href
  {https://doi.org/10.1103/PhysRevLett.126.215302} {\bibfield  {journal}
  {\bibinfo  {journal} {Phys. Rev. Lett.}\ }\textbf {\bibinfo {volume} {126}},\
  \bibinfo {pages} {215302} (\bibinfo {year} {2021})}\BibitemShut {NoStop}%
\bibitem [{\citenamefont {Wang}\ \emph {et~al.}(2023)\citenamefont {Wang},
  \citenamefont {Meng},\ and\ \citenamefont {Chen}}]{Metamaterial}%
  \BibitemOpen
  \bibfield  {author} {\bibinfo {author} {\bibfnamefont {A.}~\bibnamefont
  {Wang}}, \bibinfo {author} {\bibfnamefont {Z.}~\bibnamefont {Meng}},\ and\
  \bibinfo {author} {\bibfnamefont {C.~Q.}\ \bibnamefont {Chen}},\ }\bibfield
  {title} {\bibinfo {title} {Non-hermitian topology in static mechanical
  metamaterials},\ }\href {https://doi.org/10.1126/sciadv.adf7299} {\bibfield
  {journal} {\bibinfo  {journal} {Science Advances}\ }\textbf {\bibinfo
  {volume} {9}},\ \bibinfo {pages} {eadf7299} (\bibinfo {year} {2023})},\
  \Eprint
  {https://arxiv.org/abs/https://www.science.org/doi/pdf/10.1126/sciadv.adf7299}
  {https://www.science.org/doi/pdf/10.1126/sciadv.adf7299} \BibitemShut
  {NoStop}%
\bibitem [{\citenamefont {Liang}\ \emph {et~al.}(2022)\citenamefont {Liang},
  \citenamefont {Xie}, \citenamefont {Dong}, \citenamefont {Li}, \citenamefont
  {Li}, \citenamefont {Gadway}, \citenamefont {Yi},\ and\ \citenamefont
  {Yan}}]{Ultracold_PhysRevLett.129.070401}%
  \BibitemOpen
  \bibfield  {author} {\bibinfo {author} {\bibfnamefont {Q.}~\bibnamefont
  {Liang}}, \bibinfo {author} {\bibfnamefont {D.}~\bibnamefont {Xie}}, \bibinfo
  {author} {\bibfnamefont {Z.}~\bibnamefont {Dong}}, \bibinfo {author}
  {\bibfnamefont {H.}~\bibnamefont {Li}}, \bibinfo {author} {\bibfnamefont
  {H.}~\bibnamefont {Li}}, \bibinfo {author} {\bibfnamefont {B.}~\bibnamefont
  {Gadway}}, \bibinfo {author} {\bibfnamefont {W.}~\bibnamefont {Yi}},\ and\
  \bibinfo {author} {\bibfnamefont {B.}~\bibnamefont {Yan}},\ }\bibfield
  {title} {\bibinfo {title} {Dynamic signatures of non-hermitian skin effect
  and topology in ultracold atoms},\ }\href
  {https://doi.org/10.1103/PhysRevLett.129.070401} {\bibfield  {journal}
  {\bibinfo  {journal} {Phys. Rev. Lett.}\ }\textbf {\bibinfo {volume} {129}},\
  \bibinfo {pages} {070401} (\bibinfo {year} {2022})}\BibitemShut {NoStop}%
\bibitem [{\citenamefont {Ochkan}\ \emph {et~al.}(2024)\citenamefont {Ochkan},
  \citenamefont {Chaturvedi}, \citenamefont {K{\"o}nye}, \citenamefont
  {Veyrat}, \citenamefont {Giraud}, \citenamefont {Mailly}, \citenamefont
  {Cavanna}, \citenamefont {Gennser}, \citenamefont {Hankiewicz}, \citenamefont
  {B{\"u}chner}, \citenamefont {van~den Brink}, \citenamefont {Dufouleur},\
  and\ \citenamefont {Fulga}}]{Kiril}%
  \BibitemOpen
  \bibfield  {author} {\bibinfo {author} {\bibfnamefont {K.}~\bibnamefont
  {Ochkan}}, \bibinfo {author} {\bibfnamefont {R.}~\bibnamefont {Chaturvedi}},
  \bibinfo {author} {\bibfnamefont {V.}~\bibnamefont {K{\"o}nye}}, \bibinfo
  {author} {\bibfnamefont {L.}~\bibnamefont {Veyrat}}, \bibinfo {author}
  {\bibfnamefont {R.}~\bibnamefont {Giraud}}, \bibinfo {author} {\bibfnamefont
  {D.}~\bibnamefont {Mailly}}, \bibinfo {author} {\bibfnamefont
  {A.}~\bibnamefont {Cavanna}}, \bibinfo {author} {\bibfnamefont
  {U.}~\bibnamefont {Gennser}}, \bibinfo {author} {\bibfnamefont {E.~M.}\
  \bibnamefont {Hankiewicz}}, \bibinfo {author} {\bibfnamefont
  {B.}~\bibnamefont {B{\"u}chner}}, \bibinfo {author} {\bibfnamefont
  {J.}~\bibnamefont {van~den Brink}}, \bibinfo {author} {\bibfnamefont
  {J.}~\bibnamefont {Dufouleur}},\ and\ \bibinfo {author} {\bibfnamefont
  {I.~C.}\ \bibnamefont {Fulga}},\ }\bibfield  {title} {\bibinfo {title}
  {Non-hermitian topology in a multi-terminal quantum hall device},\ }\bibfield
   {journal} {\bibinfo  {journal} {Nature Physics}\ }\href
  {https://doi.org/10.1038/s41567-023-02337-4} {10.1038/s41567-023-02337-4}
  (\bibinfo {year} {2024})\BibitemShut {NoStop}%
\bibitem [{\citenamefont {K\"onye}\ \emph {et~al.}(2023)\citenamefont
  {K\"onye}, \citenamefont {Ochkan}, \citenamefont {Chyzhykova}, \citenamefont
  {Budich}, \citenamefont {van~den Brink}, \citenamefont {Fulga},\ and\
  \citenamefont {Dufouleur}}]{koenye2023nonhermitian}%
  \BibitemOpen
  \bibfield  {author} {\bibinfo {author} {\bibfnamefont {V.}~\bibnamefont
  {K\"onye}}, \bibinfo {author} {\bibfnamefont {K.}~\bibnamefont {Ochkan}},
  \bibinfo {author} {\bibfnamefont {A.}~\bibnamefont {Chyzhykova}}, \bibinfo
  {author} {\bibfnamefont {J.~C.}\ \bibnamefont {Budich}}, \bibinfo {author}
  {\bibfnamefont {J.}~\bibnamefont {van~den Brink}}, \bibinfo {author}
  {\bibfnamefont {I.~C.}\ \bibnamefont {Fulga}},\ and\ \bibinfo {author}
  {\bibfnamefont {J.}~\bibnamefont {Dufouleur}},\ }\href@noop {} {\bibinfo
  {title} {Non-hermitian topological ohmmeter}} (\bibinfo {year} {2023}),\
  \Eprint {https://arxiv.org/abs/2308.11367} {arXiv:2308.11367
  [cond-mat.mes-hall]} \BibitemShut {NoStop}%
\bibitem [{\citenamefont {Li}\ \emph {et~al.}(2023)\citenamefont {Li},
  \citenamefont {Tang}, \citenamefont {Duan}, \citenamefont {Xu}, \citenamefont
  {Xu}, \citenamefont {Ma},\ and\ \citenamefont {Wang}}]{Libo_Ma}%
  \BibitemOpen
  \bibfield  {author} {\bibinfo {author} {\bibfnamefont {J.}~\bibnamefont
  {Li}}, \bibinfo {author} {\bibfnamefont {M.}~\bibnamefont {Tang}}, \bibinfo
  {author} {\bibfnamefont {J.}~\bibnamefont {Duan}}, \bibinfo {author}
  {\bibfnamefont {X.}~\bibnamefont {Xu}}, \bibinfo {author} {\bibfnamefont
  {K.}~\bibnamefont {Xu}}, \bibinfo {author} {\bibfnamefont {L.}~\bibnamefont
  {Ma}},\ and\ \bibinfo {author} {\bibfnamefont {J.}~\bibnamefont {Wang}},\
  }\bibfield  {title} {\bibinfo {title} {Exceptional points in a spiral ring
  cavity for enhanced biosensing},\ }\href
  {https://doi.org/10.1109/JLT.2023.3237748} {\bibfield  {journal} {\bibinfo
  {journal} {Journal of Lightwave Technology}\ }\textbf {\bibinfo {volume}
  {41}},\ \bibinfo {pages} {2870} (\bibinfo {year} {2023})}\BibitemShut
  {NoStop}%
\bibitem [{\citenamefont {Hatano}\ and\ \citenamefont
  {Nelson}(1996)}]{Hatano-Nelson_PhysRevLett.77.570}%
  \BibitemOpen
  \bibfield  {author} {\bibinfo {author} {\bibfnamefont {N.}~\bibnamefont
  {Hatano}}\ and\ \bibinfo {author} {\bibfnamefont {D.~R.}\ \bibnamefont
  {Nelson}},\ }\bibfield  {title} {\bibinfo {title} {Localization transitions
  in non-hermitian quantum mechanics},\ }\href
  {https://doi.org/10.1103/PhysRevLett.77.570} {\bibfield  {journal} {\bibinfo
  {journal} {Phys. Rev. Lett.}\ }\textbf {\bibinfo {volume} {77}},\ \bibinfo
  {pages} {570} (\bibinfo {year} {1996})}\BibitemShut {NoStop}%
\bibitem [{\citenamefont {Bender}\ \emph
  {et~al.}(2005{\natexlab{a}})\citenamefont {Bender}, \citenamefont {Jones},\
  and\ \citenamefont {Rivers}}]{BENDER2005333}%
  \BibitemOpen
  \bibfield  {author} {\bibinfo {author} {\bibfnamefont {C.~M.}\ \bibnamefont
  {Bender}}, \bibinfo {author} {\bibfnamefont {H.}~\bibnamefont {Jones}},\ and\
  \bibinfo {author} {\bibfnamefont {R.}~\bibnamefont {Rivers}},\ }\bibfield
  {title} {\bibinfo {title} {Dual $\mathcal{PT}$-symmetric quantum field
  theories},\ }\href
  {https://doi.org/https://doi.org/10.1016/j.physletb.2005.08.087} {\bibfield
  {journal} {\bibinfo  {journal} {Physics Letters B}\ }\textbf {\bibinfo
  {volume} {625}},\ \bibinfo {pages} {333} (\bibinfo {year}
  {2005}{\natexlab{a}})}\BibitemShut {NoStop}%
\bibitem [{\citenamefont {Ashida}\ \emph {et~al.}(2016)\citenamefont {Ashida},
  \citenamefont {Furukawa},\ and\ \citenamefont {Ueda}}]{article}%
  \BibitemOpen
  \bibfield  {author} {\bibinfo {author} {\bibfnamefont {Y.}~\bibnamefont
  {Ashida}}, \bibinfo {author} {\bibfnamefont {S.}~\bibnamefont {Furukawa}},\
  and\ \bibinfo {author} {\bibfnamefont {M.}~\bibnamefont {Ueda}},\ }\bibfield
  {title} {\bibinfo {title} {Parity-time symmetric quantum critical
  phenomena},\ }\href {https://doi.org/10.1038/ncomms15791} {\bibfield
  {journal} {\bibinfo  {journal} {Nature Communications}\ }\textbf {\bibinfo
  {volume} {8}} (\bibinfo {year} {2016})}\BibitemShut {NoStop}%
\bibitem [{\citenamefont {Zhang}\ \emph {et~al.}(2022)\citenamefont {Zhang},
  \citenamefont {Denner}, \citenamefont {Bzdu\ifmmode~\check{s}\else
  \v{s}\fi{}ek}, \citenamefont {Sentef},\ and\ \citenamefont
  {Neupert}}]{Neupert_PhysRevB.106.L121102}%
  \BibitemOpen
  \bibfield  {author} {\bibinfo {author} {\bibfnamefont {S.-B.}\ \bibnamefont
  {Zhang}}, \bibinfo {author} {\bibfnamefont {M.~M.}\ \bibnamefont {Denner}},
  \bibinfo {author} {\bibfnamefont {T.~c.~v.}\ \bibnamefont
  {Bzdu\ifmmode~\check{s}\else \v{s}\fi{}ek}}, \bibinfo {author} {\bibfnamefont
  {M.~A.}\ \bibnamefont {Sentef}},\ and\ \bibinfo {author} {\bibfnamefont
  {T.}~\bibnamefont {Neupert}},\ }\bibfield  {title} {\bibinfo {title}
  {Symmetry breaking and spectral structure of the interacting hatano-nelson
  model},\ }\href {https://doi.org/10.1103/PhysRevB.106.L121102} {\bibfield
  {journal} {\bibinfo  {journal} {Phys. Rev. B}\ }\textbf {\bibinfo {volume}
  {106}},\ \bibinfo {pages} {L121102} (\bibinfo {year} {2022})}\BibitemShut
  {NoStop}%
\bibitem [{\citenamefont {{'t Hooft}}(1978)}]{THOOFT19781}%
  \BibitemOpen
  \bibfield  {author} {\bibinfo {author} {\bibfnamefont {G.}~\bibnamefont {{'t
  Hooft}}},\ }\bibfield  {title} {\bibinfo {title} {On the phase transition
  towards permanent quark confinement},\ }\href
  {https://doi.org/https://doi.org/10.1016/0550-3213(78)90153-0} {\bibfield
  {journal} {\bibinfo  {journal} {Nuclear Physics B}\ }\textbf {\bibinfo
  {volume} {138}},\ \bibinfo {pages} {1} (\bibinfo {year} {1978})}\BibitemShut
  {NoStop}%
\bibitem [{\citenamefont {Elitzur}\ \emph {et~al.}(1979)\citenamefont
  {Elitzur}, \citenamefont {Pearson},\ and\ \citenamefont
  {Shigemitsu}}]{Elitzur_PhysRevD.19.3698}%
  \BibitemOpen
  \bibfield  {author} {\bibinfo {author} {\bibfnamefont {S.}~\bibnamefont
  {Elitzur}}, \bibinfo {author} {\bibfnamefont {R.~B.}\ \bibnamefont
  {Pearson}},\ and\ \bibinfo {author} {\bibfnamefont {J.}~\bibnamefont
  {Shigemitsu}},\ }\bibfield  {title} {\bibinfo {title} {Phase structure of
  discrete abelian spin and gauge systems},\ }\href
  {https://doi.org/10.1103/PhysRevD.19.3698} {\bibfield  {journal} {\bibinfo
  {journal} {Phys. Rev. D}\ }\textbf {\bibinfo {volume} {19}},\ \bibinfo
  {pages} {3698} (\bibinfo {year} {1979})}\BibitemShut {NoStop}%
\bibitem [{\citenamefont {Jos\'e}\ \emph {et~al.}(1977)\citenamefont {Jos\'e},
  \citenamefont {Kadanoff}, \citenamefont {Kirkpatrick},\ and\ \citenamefont
  {Nelson}}]{Jose_PhysRevB.16.1217}%
  \BibitemOpen
  \bibfield  {author} {\bibinfo {author} {\bibfnamefont {J.~V.}\ \bibnamefont
  {Jos\'e}}, \bibinfo {author} {\bibfnamefont {L.~P.}\ \bibnamefont
  {Kadanoff}}, \bibinfo {author} {\bibfnamefont {S.}~\bibnamefont
  {Kirkpatrick}},\ and\ \bibinfo {author} {\bibfnamefont {D.~R.}\ \bibnamefont
  {Nelson}},\ }\bibfield  {title} {\bibinfo {title} {Renormalization, vortices,
  and symmetry-breaking perturbations in the two-dimensional planar model},\
  }\href {https://doi.org/10.1103/PhysRevB.16.1217} {\bibfield  {journal}
  {\bibinfo  {journal} {Phys. Rev. B}\ }\textbf {\bibinfo {volume} {16}},\
  \bibinfo {pages} {1217} (\bibinfo {year} {1977})}\BibitemShut {NoStop}%
\bibitem [{\citenamefont {Kadanoff}(1979)}]{KADANOFF197939}%
  \BibitemOpen
  \bibfield  {author} {\bibinfo {author} {\bibfnamefont {L.~P.}\ \bibnamefont
  {Kadanoff}},\ }\bibfield  {title} {\bibinfo {title} {Multicritical behavior
  at the kosterlitz-thouless critical point},\ }\href
  {https://doi.org/https://doi.org/10.1016/0003-4916(79)90280-X} {\bibfield
  {journal} {\bibinfo  {journal} {Annals of Physics}\ }\textbf {\bibinfo
  {volume} {120}},\ \bibinfo {pages} {39} (\bibinfo {year} {1979})}\BibitemShut
  {NoStop}%
\bibitem [{\citenamefont {Hove}\ and\ \citenamefont
  {Sudb\o{}}(2003)}]{Hove-Sudbo_PhysRevE.68.046107}%
  \BibitemOpen
  \bibfield  {author} {\bibinfo {author} {\bibfnamefont {J.}~\bibnamefont
  {Hove}}\ and\ \bibinfo {author} {\bibfnamefont {A.}~\bibnamefont
  {Sudb\o{}}},\ }\bibfield  {title} {\bibinfo {title} {Criticality versus q in
  the $(2+1)$-dimensional ${Z}_{q}$ clock model},\ }\href
  {https://doi.org/10.1103/PhysRevE.68.046107} {\bibfield  {journal} {\bibinfo
  {journal} {Phys. Rev. E}\ }\textbf {\bibinfo {volume} {68}},\ \bibinfo
  {pages} {046107} (\bibinfo {year} {2003})}\BibitemShut {NoStop}%
\bibitem [{\citenamefont {Oshikawa}(2000)}]{Oshikawa_PhysRevB.61.3430}%
  \BibitemOpen
  \bibfield  {author} {\bibinfo {author} {\bibfnamefont {M.}~\bibnamefont
  {Oshikawa}},\ }\bibfield  {title} {\bibinfo {title} {Ordered phase and
  scaling in ${Z}_{n}$ models and the three-state antiferromagnetic potts model
  in three dimensions},\ }\href {https://doi.org/10.1103/PhysRevB.61.3430}
  {\bibfield  {journal} {\bibinfo  {journal} {Phys. Rev. B}\ }\textbf {\bibinfo
  {volume} {61}},\ \bibinfo {pages} {3430} (\bibinfo {year}
  {2000})}\BibitemShut {NoStop}%
\bibitem [{\citenamefont {Pujari}\ \emph {et~al.}(2015)\citenamefont {Pujari},
  \citenamefont {Alet},\ and\ \citenamefont
  {Damle}}]{Damle_PhysRevB.91.104411}%
  \BibitemOpen
  \bibfield  {author} {\bibinfo {author} {\bibfnamefont {S.}~\bibnamefont
  {Pujari}}, \bibinfo {author} {\bibfnamefont {F.}~\bibnamefont {Alet}},\ and\
  \bibinfo {author} {\bibfnamefont {K.}~\bibnamefont {Damle}},\ }\bibfield
  {title} {\bibinfo {title} {Transitions to valence-bond solid order in a
  honeycomb lattice antiferromagnet},\ }\href
  {https://doi.org/10.1103/PhysRevB.91.104411} {\bibfield  {journal} {\bibinfo
  {journal} {Phys. Rev. B}\ }\textbf {\bibinfo {volume} {91}},\ \bibinfo
  {pages} {104411} (\bibinfo {year} {2015})}\BibitemShut {NoStop}%
\bibitem [{\citenamefont {Shao}\ \emph {et~al.}(2020)\citenamefont {Shao},
  \citenamefont {Guo},\ and\ \citenamefont
  {Sandvik}}]{Sandvik_PhysRevLett.124.080602}%
  \BibitemOpen
  \bibfield  {author} {\bibinfo {author} {\bibfnamefont {H.}~\bibnamefont
  {Shao}}, \bibinfo {author} {\bibfnamefont {W.}~\bibnamefont {Guo}},\ and\
  \bibinfo {author} {\bibfnamefont {A.~W.}\ \bibnamefont {Sandvik}},\
  }\bibfield  {title} {\bibinfo {title} {Monte carlo renormalization flows in
  the space of relevant and irrelevant operators: Application to
  three-dimensional clock models},\ }\href
  {https://doi.org/10.1103/PhysRevLett.124.080602} {\bibfield  {journal}
  {\bibinfo  {journal} {Phys. Rev. Lett.}\ }\textbf {\bibinfo {volume} {124}},\
  \bibinfo {pages} {080602} (\bibinfo {year} {2020})}\BibitemShut {NoStop}%
\bibitem [{\citenamefont {Patil}\ \emph {et~al.}(2021)\citenamefont {Patil},
  \citenamefont {Shao},\ and\ \citenamefont
  {Sandvik}}]{Sandvik_PhysRevB.103.054418}%
  \BibitemOpen
  \bibfield  {author} {\bibinfo {author} {\bibfnamefont {P.}~\bibnamefont
  {Patil}}, \bibinfo {author} {\bibfnamefont {H.}~\bibnamefont {Shao}},\ and\
  \bibinfo {author} {\bibfnamefont {A.~W.}\ \bibnamefont {Sandvik}},\
  }\bibfield  {title} {\bibinfo {title} {Unconventional u(1) to ${Z}_{q}$
  crossover in quantum and classical $q$-state clock models},\ }\href
  {https://doi.org/10.1103/PhysRevB.103.054418} {\bibfield  {journal} {\bibinfo
   {journal} {Phys. Rev. B}\ }\textbf {\bibinfo {volume} {103}},\ \bibinfo
  {pages} {054418} (\bibinfo {year} {2021})}\BibitemShut {NoStop}%
\bibitem [{\citenamefont {Zamolodchikov}(1994)}]{ZAMOLODCHIKOV1994436}%
  \BibitemOpen
  \bibfield  {author} {\bibinfo {author} {\bibfnamefont {A.}~\bibnamefont
  {Zamolodchikov}},\ }\bibfield  {title} {\bibinfo {title} {Thermodynamics of
  imaginary coupled sine-gordon. dense polymer finite-size scaling function},\
  }\href {https://doi.org/https://doi.org/10.1016/0370-2693(94)90375-1}
  {\bibfield  {journal} {\bibinfo  {journal} {Physics Letters B}\ }\textbf
  {\bibinfo {volume} {335}},\ \bibinfo {pages} {436} (\bibinfo {year}
  {1994})}\BibitemShut {NoStop}%
\bibitem [{\citenamefont {FENDLEY}\ \emph {et~al.}(1993)\citenamefont
  {FENDLEY}, \citenamefont {SALEUR},\ and\ \citenamefont
  {ZAMOLODCHIKOV}}]{Fendley}%
  \BibitemOpen
  \bibfield  {author} {\bibinfo {author} {\bibfnamefont {P.}~\bibnamefont
  {FENDLEY}}, \bibinfo {author} {\bibfnamefont {H.}~\bibnamefont {SALEUR}},\
  and\ \bibinfo {author} {\bibfnamefont {A.~B.}\ \bibnamefont
  {ZAMOLODCHIKOV}},\ }\bibfield  {title} {\bibinfo {title} {Massless flows i:
  The sine-gordon and o(n) models},\ }\href
  {https://doi.org/10.1142/S0217751X93002265} {\bibfield  {journal} {\bibinfo
  {journal} {International Journal of Modern Physics A}\ }\textbf {\bibinfo
  {volume} {08}},\ \bibinfo {pages} {5717} (\bibinfo {year}
  {1993})}\BibitemShut {NoStop}%
\bibitem [{\citenamefont {Castro-Alvaredo}\ \emph {et~al.}(2017)\citenamefont
  {Castro-Alvaredo}, \citenamefont {Doyon},\ and\ \citenamefont
  {Ravanini}}]{Castro-Alvaredo_2017}%
  \BibitemOpen
  \bibfield  {author} {\bibinfo {author} {\bibfnamefont {O.~A.}\ \bibnamefont
  {Castro-Alvaredo}}, \bibinfo {author} {\bibfnamefont {B.}~\bibnamefont
  {Doyon}},\ and\ \bibinfo {author} {\bibfnamefont {F.}~\bibnamefont
  {Ravanini}},\ }\bibfield  {title} {\bibinfo {title} {Irreversibility of the
  renormalization group flow in non-unitary quantum field theory*},\ }\href
  {https://doi.org/10.1088/1751-8121/aa8a10} {\bibfield  {journal} {\bibinfo
  {journal} {Journal of Physics A: Mathematical and Theoretical}\ }\textbf
  {\bibinfo {volume} {50}},\ \bibinfo {pages} {424002} (\bibinfo {year}
  {2017})}\BibitemShut {NoStop}%
\bibitem [{Note1()}]{Note1}%
  \BibitemOpen
  \bibinfo {note} {Note that we could equally consider the ISG as being
  $\protect \EuScript {PT}$-symmetric with $\protect \EuScript {P}\theta
  \protect \EuScript {P}^{-1}=\theta +\pi /N$, but then the $\protect \EuScript
  {PT}$ symmetry would be broken in the Lagrangian (\ref
  {Eq:Leff-0}).}\BibitemShut {Stop}%
\bibitem [{\citenamefont {Gorbenko}\ \emph {et~al.}(2018)\citenamefont
  {Gorbenko}, \citenamefont {Rychkov},\ and\ \citenamefont
  {Zan}}]{Rychkov-Walking}%
  \BibitemOpen
  \bibfield  {author} {\bibinfo {author} {\bibfnamefont {V.}~\bibnamefont
  {Gorbenko}}, \bibinfo {author} {\bibfnamefont {S.}~\bibnamefont {Rychkov}},\
  and\ \bibinfo {author} {\bibfnamefont {B.}~\bibnamefont {Zan}},\ }\bibfield
  {title} {\bibinfo {title} {Walking, weak first-order transitions, and complex
  cfts},\ }\href {https://doi.org/10.1007/JHEP10(2018)108} {\bibfield
  {journal} {\bibinfo  {journal} {Journal of High Energy Physics}\ }\textbf
  {\bibinfo {volume} {2018}},\ \bibinfo {pages} {108} (\bibinfo {year}
  {2018})}\BibitemShut {NoStop}%
\bibitem [{\citenamefont {Kaplan}\ \emph {et~al.}(2009)\citenamefont {Kaplan},
  \citenamefont {Lee}, \citenamefont {Son},\ and\ \citenamefont
  {Stephanov}}]{Son_PhysRevD.80.125005}%
  \BibitemOpen
  \bibfield  {author} {\bibinfo {author} {\bibfnamefont {D.~B.}\ \bibnamefont
  {Kaplan}}, \bibinfo {author} {\bibfnamefont {J.-W.}\ \bibnamefont {Lee}},
  \bibinfo {author} {\bibfnamefont {D.~T.}\ \bibnamefont {Son}},\ and\ \bibinfo
  {author} {\bibfnamefont {M.~A.}\ \bibnamefont {Stephanov}},\ }\bibfield
  {title} {\bibinfo {title} {Conformality lost},\ }\href
  {https://doi.org/10.1103/PhysRevD.80.125005} {\bibfield  {journal} {\bibinfo
  {journal} {Phys. Rev. D}\ }\textbf {\bibinfo {volume} {80}},\ \bibinfo
  {pages} {125005} (\bibinfo {year} {2009})}\BibitemShut {NoStop}%
\bibitem [{\citenamefont {Nogueira}\ and\ \citenamefont
  {Sudbø}(2013)}]{Nogueira_2013}%
  \BibitemOpen
  \bibfield  {author} {\bibinfo {author} {\bibfnamefont {F.~S.}\ \bibnamefont
  {Nogueira}}\ and\ \bibinfo {author} {\bibfnamefont {A.}~\bibnamefont
  {Sudbø}},\ }\bibfield  {title} {\bibinfo {title} {Deconfined quantum
  criticality and conformal phase transition in two-dimensional
  antiferromagnets},\ }\href {https://doi.org/10.1209/0295-5075/104/56004}
  {\bibfield  {journal} {\bibinfo  {journal} {Europhysics Letters}\ }\textbf
  {\bibinfo {volume} {104}},\ \bibinfo {pages} {56004} (\bibinfo {year}
  {2013})}\BibitemShut {NoStop}%
\bibitem [{\citenamefont {Nahum}\ \emph {et~al.}(2015)\citenamefont {Nahum},
  \citenamefont {Chalker}, \citenamefont {Serna}, \citenamefont {Ortu\~no},\
  and\ \citenamefont {Somoza}}]{Nahum_PhysRevX.5.041048}%
  \BibitemOpen
  \bibfield  {author} {\bibinfo {author} {\bibfnamefont {A.}~\bibnamefont
  {Nahum}}, \bibinfo {author} {\bibfnamefont {J.~T.}\ \bibnamefont {Chalker}},
  \bibinfo {author} {\bibfnamefont {P.}~\bibnamefont {Serna}}, \bibinfo
  {author} {\bibfnamefont {M.}~\bibnamefont {Ortu\~no}},\ and\ \bibinfo
  {author} {\bibfnamefont {A.~M.}\ \bibnamefont {Somoza}},\ }\bibfield  {title}
  {\bibinfo {title} {Deconfined quantum criticality, scaling violations, and
  classical loop models},\ }\href {https://doi.org/10.1103/PhysRevX.5.041048}
  {\bibfield  {journal} {\bibinfo  {journal} {Phys. Rev. X}\ }\textbf {\bibinfo
  {volume} {5}},\ \bibinfo {pages} {041048} (\bibinfo {year}
  {2015})}\BibitemShut {NoStop}%
\bibitem [{\citenamefont {Nogueira}\ \emph {et~al.}(2019)\citenamefont
  {Nogueira}, \citenamefont {van~den Brink},\ and\ \citenamefont
  {Sudb\o{}}}]{Nogueira_PhysRevD.100.085005}%
  \BibitemOpen
  \bibfield  {author} {\bibinfo {author} {\bibfnamefont {F.~S.}\ \bibnamefont
  {Nogueira}}, \bibinfo {author} {\bibfnamefont {J.}~\bibnamefont {van~den
  Brink}},\ and\ \bibinfo {author} {\bibfnamefont {A.}~\bibnamefont
  {Sudb\o{}}},\ }\bibfield  {title} {\bibinfo {title} {Conformality loss and
  quantum criticality in topological higgs electrodynamics in $2+1$
  dimensions},\ }\href {https://doi.org/10.1103/PhysRevD.100.085005} {\bibfield
   {journal} {\bibinfo  {journal} {Phys. Rev. D}\ }\textbf {\bibinfo {volume}
  {100}},\ \bibinfo {pages} {085005} (\bibinfo {year} {2019})}\BibitemShut
  {NoStop}%
\bibitem [{\citenamefont {Ma}\ and\ \citenamefont
  {Wang}(2020)}]{Ma-Wang_PhysRevB.102.020407}%
  \BibitemOpen
  \bibfield  {author} {\bibinfo {author} {\bibfnamefont {R.}~\bibnamefont
  {Ma}}\ and\ \bibinfo {author} {\bibfnamefont {C.}~\bibnamefont {Wang}},\
  }\bibfield  {title} {\bibinfo {title} {Theory of deconfined
  pseudocriticality},\ }\href {https://doi.org/10.1103/PhysRevB.102.020407}
  {\bibfield  {journal} {\bibinfo  {journal} {Phys. Rev. B}\ }\textbf {\bibinfo
  {volume} {102}},\ \bibinfo {pages} {020407} (\bibinfo {year}
  {2020})}\BibitemShut {NoStop}%
\bibitem [{\citenamefont {Ihrig}\ \emph {et~al.}(2019)\citenamefont {Ihrig},
  \citenamefont {Zerf}, \citenamefont {Marquard}, \citenamefont {Herbut},\ and\
  \citenamefont {Scherer}}]{Scherer_PhysRevB.100.134507}%
  \BibitemOpen
  \bibfield  {author} {\bibinfo {author} {\bibfnamefont {B.}~\bibnamefont
  {Ihrig}}, \bibinfo {author} {\bibfnamefont {N.}~\bibnamefont {Zerf}},
  \bibinfo {author} {\bibfnamefont {P.}~\bibnamefont {Marquard}}, \bibinfo
  {author} {\bibfnamefont {I.~F.}\ \bibnamefont {Herbut}},\ and\ \bibinfo
  {author} {\bibfnamefont {M.~M.}\ \bibnamefont {Scherer}},\ }\bibfield
  {title} {\bibinfo {title} {Abelian higgs model at four loops, fixed-point
  collision, and deconfined criticality},\ }\href
  {https://doi.org/10.1103/PhysRevB.100.134507} {\bibfield  {journal} {\bibinfo
   {journal} {Phys. Rev. B}\ }\textbf {\bibinfo {volume} {100}},\ \bibinfo
  {pages} {134507} (\bibinfo {year} {2019})}\BibitemShut {NoStop}%
\bibitem [{\citenamefont {Ma}\ and\ \citenamefont
  {He}(2019)}]{Ma-He_PhysRevB.99.195130}%
  \BibitemOpen
  \bibfield  {author} {\bibinfo {author} {\bibfnamefont {H.}~\bibnamefont
  {Ma}}\ and\ \bibinfo {author} {\bibfnamefont {Y.-C.}\ \bibnamefont {He}},\
  }\bibfield  {title} {\bibinfo {title} {Shadow of complex fixed point:
  Approximate conformality of $q>4$ potts model},\ }\href
  {https://doi.org/10.1103/PhysRevB.99.195130} {\bibfield  {journal} {\bibinfo
  {journal} {Phys. Rev. B}\ }\textbf {\bibinfo {volume} {99}},\ \bibinfo
  {pages} {195130} (\bibinfo {year} {2019})}\BibitemShut {NoStop}%
\bibitem [{\citenamefont {Burgelman}\ \emph {et~al.}(2023)\citenamefont
  {Burgelman}, \citenamefont {Devos}, \citenamefont {Vanhecke}, \citenamefont
  {Verstraete},\ and\ \citenamefont
  {Vanderstraeten}}]{Burgelman_PhysRevE.107.014117}%
  \BibitemOpen
  \bibfield  {author} {\bibinfo {author} {\bibfnamefont {L.}~\bibnamefont
  {Burgelman}}, \bibinfo {author} {\bibfnamefont {L.}~\bibnamefont {Devos}},
  \bibinfo {author} {\bibfnamefont {B.}~\bibnamefont {Vanhecke}}, \bibinfo
  {author} {\bibfnamefont {F.}~\bibnamefont {Verstraete}},\ and\ \bibinfo
  {author} {\bibfnamefont {L.}~\bibnamefont {Vanderstraeten}},\ }\bibfield
  {title} {\bibinfo {title} {Contrasting pseudocriticality in the classical
  two-dimensional heisenberg and ${\mathrm{rp}}^{2}$ models: Zero-temperature
  phase transition versus finite-temperature crossover},\ }\href
  {https://doi.org/10.1103/PhysRevE.107.014117} {\bibfield  {journal} {\bibinfo
   {journal} {Phys. Rev. E}\ }\textbf {\bibinfo {volume} {107}},\ \bibinfo
  {pages} {014117} (\bibinfo {year} {2023})}\BibitemShut {NoStop}%
\bibitem [{\citenamefont {Weber}\ and\ \citenamefont
  {Vojta}(2023)}]{Weber-Vojta_PhysRevLett.130.186701}%
  \BibitemOpen
  \bibfield  {author} {\bibinfo {author} {\bibfnamefont {M.}~\bibnamefont
  {Weber}}\ and\ \bibinfo {author} {\bibfnamefont {M.}~\bibnamefont {Vojta}},\
  }\bibfield  {title} {\bibinfo {title} {Su(2)-symmetric spin-boson model:
  Quantum criticality, fixed-point annihilation, and duality},\ }\href
  {https://doi.org/10.1103/PhysRevLett.130.186701} {\bibfield  {journal}
  {\bibinfo  {journal} {Phys. Rev. Lett.}\ }\textbf {\bibinfo {volume} {130}},\
  \bibinfo {pages} {186701} (\bibinfo {year} {2023})}\BibitemShut {NoStop}%
\bibitem [{\citenamefont {Hawashin}\ \emph {et~al.}(2023)\citenamefont
  {Hawashin}, \citenamefont {Eichhorn}, \citenamefont {Janssen}, \citenamefont
  {Scherer},\ and\ \citenamefont {Ray}}]{hawashin2023nordicwalking}%
  \BibitemOpen
  \bibfield  {author} {\bibinfo {author} {\bibfnamefont {B.}~\bibnamefont
  {Hawashin}}, \bibinfo {author} {\bibfnamefont {A.}~\bibnamefont {Eichhorn}},
  \bibinfo {author} {\bibfnamefont {L.}~\bibnamefont {Janssen}}, \bibinfo
  {author} {\bibfnamefont {M.~M.}\ \bibnamefont {Scherer}},\ and\ \bibinfo
  {author} {\bibfnamefont {S.}~\bibnamefont {Ray}},\ }\href@noop {} {\bibinfo
  {title} {The nordic-walking mechanism and its explanation of deconfined
  pseudocriticality from wess-zumino-witten theory}} (\bibinfo {year} {2023}),\
  \Eprint {https://arxiv.org/abs/2312.11614} {arXiv:2312.11614
  [cond-mat.str-el]} \BibitemShut {NoStop}%
\bibitem [{\citenamefont {Weber}(2024)}]{weber2024tunable}%
  \BibitemOpen
  \bibfield  {author} {\bibinfo {author} {\bibfnamefont {M.}~\bibnamefont
  {Weber}},\ }\href@noop {} {\bibinfo {title} {Tunable quantum criticality and
  pseudocriticality across the fixed-point annihilation in the anisotropic
  spin-boson model}} (\bibinfo {year} {2024}),\ \Eprint
  {https://arxiv.org/abs/2403.02400} {arXiv:2403.02400 [cond-mat.str-el]}
  \BibitemShut {NoStop}%
\bibitem [{\citenamefont {Zhou}\ \emph {et~al.}(2024)\citenamefont {Zhou},
  \citenamefont {Hu}, \citenamefont {Zhu},\ and\ \citenamefont
  {He}}]{zhou2024mathrmso5}%
  \BibitemOpen
  \bibfield  {author} {\bibinfo {author} {\bibfnamefont {Z.}~\bibnamefont
  {Zhou}}, \bibinfo {author} {\bibfnamefont {L.}~\bibnamefont {Hu}}, \bibinfo
  {author} {\bibfnamefont {W.}~\bibnamefont {Zhu}},\ and\ \bibinfo {author}
  {\bibfnamefont {Y.-C.}\ \bibnamefont {He}},\ }\href@noop {} {\bibinfo {title}
  {The $\mathrm{SO}(5)$ deconfined phase transition under the fuzzy sphere
  microscope: Approximate conformal symmetry, pseudo-criticality, and operator
  spectrum}} (\bibinfo {year} {2024}),\ \Eprint
  {https://arxiv.org/abs/2306.16435} {arXiv:2306.16435 [cond-mat.str-el]}
  \BibitemShut {NoStop}%
\bibitem [{\citenamefont {Jacobsen}\ and\ \citenamefont
  {Wiese}(2024)}]{jacobsen2024lattice}%
  \BibitemOpen
  \bibfield  {author} {\bibinfo {author} {\bibfnamefont {J.~L.}\ \bibnamefont
  {Jacobsen}}\ and\ \bibinfo {author} {\bibfnamefont {K.~J.}\ \bibnamefont
  {Wiese}},\ }\href@noop {} {\bibinfo {title} {Lattice realization of complex
  cfts: Two-dimensional potts model with $q>4$ states}} (\bibinfo {year}
  {2024}),\ \Eprint {https://arxiv.org/abs/2402.10732} {arXiv:2402.10732
  [hep-th]} \BibitemShut {NoStop}%
\bibitem [{\citenamefont {Tang}\ \emph {et~al.}(2024)\citenamefont {Tang},
  \citenamefont {Ma}, \citenamefont {Tang}, \citenamefont {He},\ and\
  \citenamefont {Zhu}}]{tang2024reclaiming}%
  \BibitemOpen
  \bibfield  {author} {\bibinfo {author} {\bibfnamefont {Y.}~\bibnamefont
  {Tang}}, \bibinfo {author} {\bibfnamefont {H.}~\bibnamefont {Ma}}, \bibinfo
  {author} {\bibfnamefont {Q.}~\bibnamefont {Tang}}, \bibinfo {author}
  {\bibfnamefont {Y.-C.}\ \bibnamefont {He}},\ and\ \bibinfo {author}
  {\bibfnamefont {W.}~\bibnamefont {Zhu}},\ }\href@noop {} {\bibinfo {title}
  {Reclaiming the lost conformality in a non-hermitian quantum 5-state potts
  model}} (\bibinfo {year} {2024}),\ \Eprint {https://arxiv.org/abs/2403.00852}
  {arXiv:2403.00852 [cond-mat.stat-mech]} \BibitemShut {NoStop}%
\bibitem [{\citenamefont {Ikhlef}\ \emph {et~al.}(2016)\citenamefont {Ikhlef},
  \citenamefont {Jacobsen},\ and\ \citenamefont
  {Saleur}}]{Saleur_PhysRevLett.116.130601}%
  \BibitemOpen
  \bibfield  {author} {\bibinfo {author} {\bibfnamefont {Y.}~\bibnamefont
  {Ikhlef}}, \bibinfo {author} {\bibfnamefont {J.~L.}\ \bibnamefont
  {Jacobsen}},\ and\ \bibinfo {author} {\bibfnamefont {H.}~\bibnamefont
  {Saleur}},\ }\bibfield  {title} {\bibinfo {title} {Three-point functions in
  $c\ensuremath{\le}1$ liouville theory and conformal loop ensembles},\ }\href
  {https://doi.org/10.1103/PhysRevLett.116.130601} {\bibfield  {journal}
  {\bibinfo  {journal} {Phys. Rev. Lett.}\ }\textbf {\bibinfo {volume} {116}},\
  \bibinfo {pages} {130601} (\bibinfo {year} {2016})}\BibitemShut {NoStop}%
\bibitem [{\citenamefont {CARRUTHERS}\ and\ \citenamefont
  {NIETO}(1968)}]{Carruthers-Nieto_RevModPhys.40.411}%
  \BibitemOpen
  \bibfield  {author} {\bibinfo {author} {\bibfnamefont {P.}~\bibnamefont
  {CARRUTHERS}}\ and\ \bibinfo {author} {\bibfnamefont {M.~M.}\ \bibnamefont
  {NIETO}},\ }\bibfield  {title} {\bibinfo {title} {Phase and angle variables
  in quantum mechanics},\ }\href {https://doi.org/10.1103/RevModPhys.40.411}
  {\bibfield  {journal} {\bibinfo  {journal} {Rev. Mod. Phys.}\ }\textbf
  {\bibinfo {volume} {40}},\ \bibinfo {pages} {411} (\bibinfo {year}
  {1968})}\BibitemShut {NoStop}%
\bibitem [{\citenamefont {Pegg}\ and\ \citenamefont
  {Barnett}(1988)}]{DTPegg_1988}%
  \BibitemOpen
  \bibfield  {author} {\bibinfo {author} {\bibfnamefont {D.~T.}\ \bibnamefont
  {Pegg}}\ and\ \bibinfo {author} {\bibfnamefont {S.~M.}\ \bibnamefont
  {Barnett}},\ }\bibfield  {title} {\bibinfo {title} {Unitary phase operator in
  quantum mechanics},\ }\href {https://doi.org/10.1209/0295-5075/6/6/002}
  {\bibfield  {journal} {\bibinfo  {journal} {Europhysics Letters}\ }\textbf
  {\bibinfo {volume} {6}},\ \bibinfo {pages} {483} (\bibinfo {year}
  {1988})}\BibitemShut {NoStop}%
\bibitem [{\citenamefont {Bender}\ \emph
  {et~al.}(2005{\natexlab{b}})\citenamefont {Bender}, \citenamefont {Brandt},
  \citenamefont {Chen}, \citenamefont {Wang},\ and\ \citenamefont
  {Qinghai}}]{Bender_PhysRevD.71.025014}%
  \BibitemOpen
  \bibfield  {author} {\bibinfo {author} {\bibfnamefont {C.~M.}\ \bibnamefont
  {Bender}}, \bibinfo {author} {\bibfnamefont {S.~F.}\ \bibnamefont {Brandt}},
  \bibinfo {author} {\bibfnamefont {J.-H.}\ \bibnamefont {Chen}}, \bibinfo
  {author} {\bibnamefont {Wang}},\ and\ \bibinfo {author} {\bibnamefont
  {Qinghai}},\ }\bibfield  {title} {\bibinfo {title} {Ghost busting:
  $\mathcal{P}\mathcal{T}$-symmetric interpretation of the lee model},\ }\href
  {https://doi.org/10.1103/PhysRevD.71.025014} {\bibfield  {journal} {\bibinfo
  {journal} {Phys. Rev. D}\ }\textbf {\bibinfo {volume} {71}},\ \bibinfo
  {pages} {025014} (\bibinfo {year} {2005}{\natexlab{b}})}\BibitemShut
  {NoStop}%
\bibitem [{\citenamefont {Gati}\ and\ \citenamefont
  {Oberthaler}(2007)}]{Oberthaler_2007}%
  \BibitemOpen
  \bibfield  {author} {\bibinfo {author} {\bibfnamefont {R.}~\bibnamefont
  {Gati}}\ and\ \bibinfo {author} {\bibfnamefont {M.~K.}\ \bibnamefont
  {Oberthaler}},\ }\bibfield  {title} {\bibinfo {title} {A bosonic josephson
  junction},\ }\href {https://doi.org/10.1088/0953-4075/40/10/R01} {\bibfield
  {journal} {\bibinfo  {journal} {Journal of Physics B: Atomic, Molecular and
  Optical Physics}\ }\textbf {\bibinfo {volume} {40}},\ \bibinfo {pages} {R61}
  (\bibinfo {year} {2007})}\BibitemShut {NoStop}%
\bibitem [{\citenamefont {Fisher}\ \emph {et~al.}(1989)\citenamefont {Fisher},
  \citenamefont {Weichman}, \citenamefont {Grinstein},\ and\ \citenamefont
  {Fisher}}]{Bose-Hubbard_PhysRevB.40.546}%
  \BibitemOpen
  \bibfield  {author} {\bibinfo {author} {\bibfnamefont {M.~P.~A.}\
  \bibnamefont {Fisher}}, \bibinfo {author} {\bibfnamefont {P.~B.}\
  \bibnamefont {Weichman}}, \bibinfo {author} {\bibfnamefont {G.}~\bibnamefont
  {Grinstein}},\ and\ \bibinfo {author} {\bibfnamefont {D.~S.}\ \bibnamefont
  {Fisher}},\ }\bibfield  {title} {\bibinfo {title} {Boson localization and the
  superfluid-insulator transition},\ }\href
  {https://doi.org/10.1103/PhysRevB.40.546} {\bibfield  {journal} {\bibinfo
  {journal} {Phys. Rev. B}\ }\textbf {\bibinfo {volume} {40}},\ \bibinfo
  {pages} {546} (\bibinfo {year} {1989})}\BibitemShut {NoStop}%
\bibitem [{\citenamefont {Park}\ \emph {et~al.}(2022)\citenamefont {Park},
  \citenamefont {Cao}, \citenamefont {Xia}, \citenamefont {Sun}, \citenamefont
  {Watanabe}, \citenamefont {Taniguchi},\ and\ \citenamefont
  {Jarillo-Herrero}}]{Jarillo-Herrero-NatMat}%
  \BibitemOpen
  \bibfield  {author} {\bibinfo {author} {\bibfnamefont {J.~M.}\ \bibnamefont
  {Park}}, \bibinfo {author} {\bibfnamefont {Y.}~\bibnamefont {Cao}}, \bibinfo
  {author} {\bibfnamefont {L.-Q.}\ \bibnamefont {Xia}}, \bibinfo {author}
  {\bibfnamefont {S.}~\bibnamefont {Sun}}, \bibinfo {author} {\bibfnamefont
  {K.}~\bibnamefont {Watanabe}}, \bibinfo {author} {\bibfnamefont
  {T.}~\bibnamefont {Taniguchi}},\ and\ \bibinfo {author} {\bibfnamefont
  {P.}~\bibnamefont {Jarillo-Herrero}},\ }\bibfield  {title} {\bibinfo {title}
  {Robust superconductivity in magic-angle multilayer graphene family},\ }\href
  {https://doi.org/10.1038/s41563-022-01287-1} {\bibfield  {journal} {\bibinfo
  {journal} {Nature Materials}\ }\textbf {\bibinfo {volume} {21}},\ \bibinfo
  {pages} {877} (\bibinfo {year} {2022})}\BibitemShut {NoStop}%
\bibitem [{\citenamefont {Zhao}\ \emph {et~al.}(2023)\citenamefont {Zhao},
  \citenamefont {Cui}, \citenamefont {Volkov}, \citenamefont {Yoo},
  \citenamefont {Lee}, \citenamefont {Gardener}, \citenamefont {Akey},
  \citenamefont {Engelke}, \citenamefont {Ronen}, \citenamefont {Zhong},
  \citenamefont {Gu}, \citenamefont {Plugge}, \citenamefont {Tummuru},
  \citenamefont {Kim}, \citenamefont {Franz}, \citenamefont {Pixley},
  \citenamefont {Poccia},\ and\ \citenamefont {Kim}}]{Poccia-Kim_Science}%
  \BibitemOpen
  \bibfield  {author} {\bibinfo {author} {\bibfnamefont {S.~Y.~F.}\
  \bibnamefont {Zhao}}, \bibinfo {author} {\bibfnamefont {X.}~\bibnamefont
  {Cui}}, \bibinfo {author} {\bibfnamefont {P.~A.}\ \bibnamefont {Volkov}},
  \bibinfo {author} {\bibfnamefont {H.}~\bibnamefont {Yoo}}, \bibinfo {author}
  {\bibfnamefont {S.}~\bibnamefont {Lee}}, \bibinfo {author} {\bibfnamefont
  {J.~A.}\ \bibnamefont {Gardener}}, \bibinfo {author} {\bibfnamefont {A.~J.}\
  \bibnamefont {Akey}}, \bibinfo {author} {\bibfnamefont {R.}~\bibnamefont
  {Engelke}}, \bibinfo {author} {\bibfnamefont {Y.}~\bibnamefont {Ronen}},
  \bibinfo {author} {\bibfnamefont {R.}~\bibnamefont {Zhong}}, \bibinfo
  {author} {\bibfnamefont {G.}~\bibnamefont {Gu}}, \bibinfo {author}
  {\bibfnamefont {S.}~\bibnamefont {Plugge}}, \bibinfo {author} {\bibfnamefont
  {T.}~\bibnamefont {Tummuru}}, \bibinfo {author} {\bibfnamefont
  {M.}~\bibnamefont {Kim}}, \bibinfo {author} {\bibfnamefont {M.}~\bibnamefont
  {Franz}}, \bibinfo {author} {\bibfnamefont {J.~H.}\ \bibnamefont {Pixley}},
  \bibinfo {author} {\bibfnamefont {N.}~\bibnamefont {Poccia}},\ and\ \bibinfo
  {author} {\bibfnamefont {P.}~\bibnamefont {Kim}},\ }\bibfield  {title}
  {\bibinfo {title} {Time-reversal symmetry breaking superconductivity between
  twisted cuprate superconductors},\ }\href
  {https://doi.org/10.1126/science.abl8371} {\bibfield  {journal} {\bibinfo
  {journal} {Science}\ }\textbf {\bibinfo {volume} {382}},\ \bibinfo {pages}
  {1422} (\bibinfo {year} {2023})},\ \Eprint
  {https://arxiv.org/abs/https://www.science.org/doi/pdf/10.1126/science.abl8371}
  {https://www.science.org/doi/pdf/10.1126/science.abl8371} \BibitemShut
  {NoStop}%
\bibitem [{\citenamefont {Fr{\"o}hlich}\ and\ \citenamefont
  {Spencer}(1981)}]{frohlich1981statistical}%
  \BibitemOpen
  \bibfield  {author} {\bibinfo {author} {\bibfnamefont {J.}~\bibnamefont
  {Fr{\"o}hlich}}\ and\ \bibinfo {author} {\bibfnamefont {T.}~\bibnamefont
  {Spencer}},\ }\bibfield  {title} {\bibinfo {title} {On the statistical
  mechanics of classical coulomb and dipole gases},\ }\href
  {https://doi.org/10.1007/BF01011379} {\bibfield  {journal} {\bibinfo
  {journal} {Journal of Statistical Physics}\ }\textbf {\bibinfo {volume}
  {24}},\ \bibinfo {pages} {617} (\bibinfo {year} {1981})}\BibitemShut
  {NoStop}%
\bibitem [{\citenamefont {Nelson}\ and\ \citenamefont
  {Kosterlitz}(1977)}]{Nelson-Kosterlitz1977}%
  \BibitemOpen
  \bibfield  {author} {\bibinfo {author} {\bibfnamefont {D.~R.}\ \bibnamefont
  {Nelson}}\ and\ \bibinfo {author} {\bibfnamefont {J.~M.}\ \bibnamefont
  {Kosterlitz}},\ }\bibfield  {title} {\bibinfo {title} {Universal jump in the
  superfluid density of two-dimensional superfluids},\ }\href
  {https://doi.org/10.1103/PhysRevLett.39.1201} {\bibfield  {journal} {\bibinfo
   {journal} {Phys. Rev. Lett.}\ }\textbf {\bibinfo {volume} {39}},\ \bibinfo
  {pages} {1201} (\bibinfo {year} {1977})}\BibitemShut {NoStop}%
\bibitem [{\citenamefont {Zinn-Justin}(2021)}]{zinn2021quantum}%
  \BibitemOpen
  \bibfield  {author} {\bibinfo {author} {\bibfnamefont {J.}~\bibnamefont
  {Zinn-Justin}},\ }\href {https://books.google.de/books?id=ioskEAAAQBAJ}
  {\emph {\bibinfo {title} {Quantum Field Theory and Critical Phenomena: Fifth
  Edition}}},\ International series of monographs on physics\ (\bibinfo
  {publisher} {Oxford University Press},\ \bibinfo {year} {2021})\BibitemShut
  {NoStop}%
\bibitem [{\citenamefont {Herbut}(2016)}]{Herbut_PhysRevD.94.025036}%
  \BibitemOpen
  \bibfield  {author} {\bibinfo {author} {\bibfnamefont {I.~F.}\ \bibnamefont
  {Herbut}},\ }\bibfield  {title} {\bibinfo {title} {Chiral symmetry breaking
  in three-dimensional quantum electrodynamics as fixed point annihilation},\
  }\href {https://doi.org/10.1103/PhysRevD.94.025036} {\bibfield  {journal}
  {\bibinfo  {journal} {Phys. Rev. D}\ }\textbf {\bibinfo {volume} {94}},\
  \bibinfo {pages} {025036} (\bibinfo {year} {2016})}\BibitemShut {NoStop}%
\bibitem [{\citenamefont {Miransky}\ and\ \citenamefont
  {Yamawaki}(1997)}]{Miransky_PhysRevD.55.5051}%
  \BibitemOpen
  \bibfield  {author} {\bibinfo {author} {\bibfnamefont {V.~A.}\ \bibnamefont
  {Miransky}}\ and\ \bibinfo {author} {\bibfnamefont {K.}~\bibnamefont
  {Yamawaki}},\ }\bibfield  {title} {\bibinfo {title} {Conformal phase
  transition in gauge theories},\ }\href
  {https://doi.org/10.1103/PhysRevD.55.5051} {\bibfield  {journal} {\bibinfo
  {journal} {Phys. Rev. D}\ }\textbf {\bibinfo {volume} {55}},\ \bibinfo
  {pages} {5051} (\bibinfo {year} {1997})}\BibitemShut {NoStop}%
\bibitem [{\citenamefont {Amit}\ and\ \citenamefont
  {Peliti}(1982)}]{AMIT1982207}%
  \BibitemOpen
  \bibfield  {author} {\bibinfo {author} {\bibfnamefont {D.~J.}\ \bibnamefont
  {Amit}}\ and\ \bibinfo {author} {\bibfnamefont {L.}~\bibnamefont {Peliti}},\
  }\bibfield  {title} {\bibinfo {title} {On dangerous irrelevant operators},\
  }\href {https://doi.org/https://doi.org/10.1016/0003-4916(82)90159-2}
  {\bibfield  {journal} {\bibinfo  {journal} {Annals of Physics}\ }\textbf
  {\bibinfo {volume} {140}},\ \bibinfo {pages} {207} (\bibinfo {year}
  {1982})}\BibitemShut {NoStop}%
\bibitem [{\citenamefont {B{\o}ttcher}\ \emph {et~al.}(2018)\citenamefont
  {B{\o}ttcher}, \citenamefont {Nichele}, \citenamefont {Kjaergaard},
  \citenamefont {Suominen}, \citenamefont {Shabani}, \citenamefont
  {Palmstr{\o}m},\ and\ \citenamefont {Marcus}}]{Marcus}%
  \BibitemOpen
  \bibfield  {author} {\bibinfo {author} {\bibfnamefont {C.~G.~L.}\
  \bibnamefont {B{\o}ttcher}}, \bibinfo {author} {\bibfnamefont
  {F.}~\bibnamefont {Nichele}}, \bibinfo {author} {\bibfnamefont
  {M.}~\bibnamefont {Kjaergaard}}, \bibinfo {author} {\bibfnamefont {H.~J.}\
  \bibnamefont {Suominen}}, \bibinfo {author} {\bibfnamefont {J.}~\bibnamefont
  {Shabani}}, \bibinfo {author} {\bibfnamefont {C.~J.}\ \bibnamefont
  {Palmstr{\o}m}},\ and\ \bibinfo {author} {\bibfnamefont {C.~M.}\ \bibnamefont
  {Marcus}},\ }\bibfield  {title} {\bibinfo {title} {Superconducting,
  insulating and anomalous metallic regimes in a gated two-dimensional
  semiconductor--superconductor array},\ }\href
  {https://doi.org/10.1038/s41567-018-0259-9} {\bibfield  {journal} {\bibinfo
  {journal} {Nature Physics}\ }\textbf {\bibinfo {volume} {14}},\ \bibinfo
  {pages} {1138} (\bibinfo {year} {2018})}\BibitemShut {NoStop}%
\bibitem [{\citenamefont {Nogueira}\ and\ \citenamefont
  {Kleinert}(2008)}]{Nogueira-Kleinert_PhysRevB.77.045107}%
  \BibitemOpen
  \bibfield  {author} {\bibinfo {author} {\bibfnamefont {F.~S.}\ \bibnamefont
  {Nogueira}}\ and\ \bibinfo {author} {\bibfnamefont {H.}~\bibnamefont
  {Kleinert}},\ }\bibfield  {title} {\bibinfo {title} {Compact quantum
  electrodynamics in $2+1$ dimensions and spinon deconfinement: A
  renormalization group analysis},\ }\href
  {https://doi.org/10.1103/PhysRevB.77.045107} {\bibfield  {journal} {\bibinfo
  {journal} {Phys. Rev. B}\ }\textbf {\bibinfo {volume} {77}},\ \bibinfo
  {pages} {045107} (\bibinfo {year} {2008})}\BibitemShut {NoStop}%
\bibitem [{\citenamefont {Kosterlitz}(1977)}]{Kosterlitz_1977}%
  \BibitemOpen
  \bibfield  {author} {\bibinfo {author} {\bibfnamefont {J.~M.}\ \bibnamefont
  {Kosterlitz}},\ }\bibfield  {title} {\bibinfo {title} {The d-dimensional
  coulomb gas and the roughening transition},\ }\href
  {https://doi.org/10.1088/0022-3719/10/19/011} {\bibfield  {journal} {\bibinfo
   {journal} {Journal of Physics C: Solid State Physics}\ }\textbf {\bibinfo
  {volume} {10}},\ \bibinfo {pages} {3753} (\bibinfo {year}
  {1977})}\BibitemShut {NoStop}%
\bibitem [{\citenamefont {Kleinert}\ \emph {et~al.}(2003)\citenamefont
  {Kleinert}, \citenamefont {Nogueira},\ and\ \citenamefont
  {Sudbø}}]{KLEINERT2003361}%
  \BibitemOpen
  \bibfield  {author} {\bibinfo {author} {\bibfnamefont {H.}~\bibnamefont
  {Kleinert}}, \bibinfo {author} {\bibfnamefont {F.~S.}\ \bibnamefont
  {Nogueira}},\ and\ \bibinfo {author} {\bibfnamefont {A.}~\bibnamefont
  {Sudbø}},\ }\bibfield  {title} {\bibinfo {title} {Kosterlitz–thouless-like
  deconfinement mechanism in the (2+1)-dimensional abelian higgs model},\
  }\href {https://doi.org/https://doi.org/10.1016/S0550-3213(03)00453-X}
  {\bibfield  {journal} {\bibinfo  {journal} {Nuclear Physics B}\ }\textbf
  {\bibinfo {volume} {666}},\ \bibinfo {pages} {361} (\bibinfo {year}
  {2003})}\BibitemShut {NoStop}%
\bibitem [{\citenamefont {Peskin}(1978)}]{Peskin1978}%
  \BibitemOpen
  \bibfield  {author} {\bibinfo {author} {\bibfnamefont {M.~E.}\ \bibnamefont
  {Peskin}},\ }\bibfield  {title} {\bibinfo {title} {{Mandelstam-'t Hooft
  duality in abelian lattice models}},\ }\href
  {https://doi.org/10.1016/0003-4916(78)90252-X} {\bibfield  {journal}
  {\bibinfo  {journal} {Ann. Phys. (N. Y).}\ }\textbf {\bibinfo {volume}
  {113}},\ \bibinfo {pages} {122} (\bibinfo {year} {1978})}\BibitemShut
  {NoStop}%
\bibitem [{\citenamefont {Dasgupta}\ and\ \citenamefont
  {Halperin}(1981)}]{Dasgupta-Halperin_PhysRevLett.47.1556}%
  \BibitemOpen
  \bibfield  {author} {\bibinfo {author} {\bibfnamefont {C.}~\bibnamefont
  {Dasgupta}}\ and\ \bibinfo {author} {\bibfnamefont {B.~I.}\ \bibnamefont
  {Halperin}},\ }\bibfield  {title} {\bibinfo {title} {Phase transition in a
  lattice model of superconductivity},\ }\href
  {https://doi.org/10.1103/PhysRevLett.47.1556} {\bibfield  {journal} {\bibinfo
   {journal} {Phys. Rev. Lett.}\ }\textbf {\bibinfo {volume} {47}},\ \bibinfo
  {pages} {1556} (\bibinfo {year} {1981})}\BibitemShut {NoStop}%
\bibitem [{\citenamefont {Kleinert}(1989)}]{kleinert1989gauge}%
  \BibitemOpen
  \bibfield  {author} {\bibinfo {author} {\bibfnamefont {H.}~\bibnamefont
  {Kleinert}},\ }\href@noop {} {\emph {\bibinfo {title} {Gauge Fields in
  Condensed Matter: Vol. 1: Superflow and Vortex Lines (Disorder Fields, Phase
  Transitions) Vol. 2: Stresses and Defects (Differential Geometry, Crystal
  Melting)}}}\ (\bibinfo  {publisher} {World Scientific},\ \bibinfo {year}
  {1989})\BibitemShut {NoStop}%
\bibitem [{\citenamefont {Nelson}\ and\ \citenamefont
  {Fisher}(1977)}]{Nelson-Fisher_PhysRevB.16.4945}%
  \BibitemOpen
  \bibfield  {author} {\bibinfo {author} {\bibfnamefont {D.~R.}\ \bibnamefont
  {Nelson}}\ and\ \bibinfo {author} {\bibfnamefont {D.~S.}\ \bibnamefont
  {Fisher}},\ }\bibfield  {title} {\bibinfo {title} {Dynamics of classical
  $\mathrm{XY}$ spins in one and two dimensions},\ }\href
  {https://doi.org/10.1103/PhysRevB.16.4945} {\bibfield  {journal} {\bibinfo
  {journal} {Phys. Rev. B}\ }\textbf {\bibinfo {volume} {16}},\ \bibinfo
  {pages} {4945} (\bibinfo {year} {1977})}\BibitemShut {NoStop}%
\bibitem [{\citenamefont {Cardy}\ and\ \citenamefont
  {Hamber}(1980)}]{Cardy-Hamber_PhysRevLett.45.499}%
  \BibitemOpen
  \bibfield  {author} {\bibinfo {author} {\bibfnamefont {J.~L.}\ \bibnamefont
  {Cardy}}\ and\ \bibinfo {author} {\bibfnamefont {H.~W.}\ \bibnamefont
  {Hamber}},\ }\bibfield  {title} {\bibinfo {title} {$o(n)$ heisenberg model
  close to $n=d=2$},\ }\href {https://doi.org/10.1103/PhysRevLett.45.499}
  {\bibfield  {journal} {\bibinfo  {journal} {Phys. Rev. Lett.}\ }\textbf
  {\bibinfo {volume} {45}},\ \bibinfo {pages} {499} (\bibinfo {year}
  {1980})}\BibitemShut {NoStop}%
\bibitem [{\citenamefont {Shenoy}(1989)}]{Shenoy_PhysRevB.40.5056}%
  \BibitemOpen
  \bibfield  {author} {\bibinfo {author} {\bibfnamefont {S.~R.}\ \bibnamefont
  {Shenoy}},\ }\bibfield  {title} {\bibinfo {title} {Vortex-loop scaling in the
  three-dimensional xy ferromagnet},\ }\href
  {https://doi.org/10.1103/PhysRevB.40.5056} {\bibfield  {journal} {\bibinfo
  {journal} {Phys. Rev. B}\ }\textbf {\bibinfo {volume} {40}},\ \bibinfo
  {pages} {5056} (\bibinfo {year} {1989})}\BibitemShut {NoStop}%
\end{thebibliography}%

\end{document}